\documentclass[12pt,letterpaper]{article}
\pdfoutput=1

\usepackage{jheppub}
\usepackage{amsmath}
\usepackage{amssymb}
\usepackage{xcolor}
\usepackage{graphicx}
\usepackage{subfig}
\usepackage{bbold}
\usepackage{tabularx}
\usepackage[makeroom]{cancel}
\usepackage{multirow}
\usepackage{slashed}
\usepackage{hyperref}
\usepackage{mathtools}
\usepackage{dirtytalk}
\usepackage{tabularx}
\usepackage{rotating}
\usepackage{nicematrix}
\usepackage{caption}
\usepackage{comment}
\usepackage{array}

\newcommand{\Mpl}{M_\mathrm{Pl}}

\title{De Sitter space constraints on brane tensions and couplings}

\author[a,b]{Saquib Hassan,}

\author[a]{Georges Obied,}

\author[a]{and John March-Russell}

\emailAdd{saquib.hassan@physics.ox.ac.uk} 

\emailAdd{georges.obied@physics.ox.ac.uk}

\emailAdd{john.march-russell@physics.ox.ac.uk}

\affiliation[a]{Rudolf Peierls Centre for Theoretical Physics, University of Oxford, Oxford, OX1 3PU, United Kingdom}
\affiliation[b]{Christ Church, University of Oxford, Oxford, OX1 1DP, United Kingdom}

\date{\today}

\newcommand{\pd}{\partial}
\newcommand{\MPl}{M_{\rm Pl}}

\definecolor{darkgreen}{RGB}{50,150,0}
\newcommand{\go}[1]{\textcolor{darkgreen}{GO: #1}}

\definecolor{darkred}{RGB}{150,0,50}

\definecolor{darkblue}{RGB}{0,50,150}

\abstract{
We argue for the existence of bounds on the tensions of $p$-branes in de Sitter space in terms of the Hubble rate and the strength of a class of Chern-Simons-like couplings. The world-volume couplings involve Abelian 1-form gauge fields in the bulk and possibly field strengths intrinsic to the brane. In many cases these couplings are the D-brane Chern-Simons terms present in string theory, while in other cases they are the interactions of axion domain walls with $U(1)$ fields. Our arguments use the same logic and assumptions as the recent Festina Lente proposal (thus utilizing the properties of Nariai de Sitter black holes) and generalize it to extended objects, thereby providing a bottom-up set of constraints independent of any particular UV completion. We compare these bounds to the properties of (wrapped) D-branes in Type II string theory in the weak coupling limit, under the assumption that these properties are not modified significantly in de Sitter constructions. We find that all constraints are satisfied by D-branes, providing further evidence for the Festina Lente conjecture.  For the particular case of 2-branes with Chern-Simons interactions we obtain a bound, which however can be evaded if the theory contains a light axion. Similarly, we find the bounds do not apply to axion domain walls due to the presence of the axion.
}

\begin{document}

\maketitle

\section{Introduction}
\label{sec:introduction}

In recent years, there has been a lot of progress in understanding general properties of the landscape of quantum gravity vacua at low-energies. Identifying the conditions that can render a low-energy theory inconsistent in the presence of gravity has evolved into the extensive Swampland program~\cite{Vafa:2005ui,Ooguri:2006in} (for reviews of the current status of the program see~\cite{Palti:2019pca,grana2021swampland,vanBeest:2021lhn,Agmon:2022thq}).
These conditions are often conjectured from the top-down based on examples (or lack thereof) in string theory and hence rely on an ultra-violet (UV) description of gravity. On the other hand, many (but not all) of the Swampland criteria can be motivated directly from the bottom-up using physical principles such as unitarity (e.g.~\cite{Kim:2019vuc}), absence of anomalies (e.g.~\cite{Adams:2010zy,Montero:2020icj}), holography (e.g.~\cite{Montero:2018fns,Harlow:2018jwu,Harlow:2018tng}) or black hole physics (e.g.~\cite{Susskind:1995da,Arkani-Hamed:2006emk}). This extra structure gives strong evidence in favour of these statements and allows one to use them independent of the existence of a weakly coupled string description. In particular, these arguments rely solely on well-known properties of the low-energy theory that we expect should hold in any realistic theory of gravity. 

These bottom-up arguments can play an especially important role when trying to understand properties of de Sitter (dS) or quasi-dS\footnote{By quasi-dS, we mean accelerating spacetimes where the Hubble rate varies slowly compared to the time-scale of the expansion itself, i.e. $\dot{H} \ll H^2$. In the context of Festina Lente, a similar condition has also been formulated in terms of the potential of a rolling scalar field in~\cite{Montero:2021otb}.} quantum gravity where the lack of parametric control can make it difficult to establish top-down statements. The Festina Lente (FL) bound \cite{Montero:2019ekk,Montero:2021otb} is one such example that can be derived using Einstein-Maxwell theory in dS space without reference to its UV origins. The original statement of FL\footnote{We will review the FL bound, along with potential caveats, in detail in section~\ref{sec:FLReview} but for now we proceed with a high-level discussion.} is a lower bound on the masses of particles in (quasi-)dS space that are charged under a $U(1)$ gauge group:
\begin{align}
\label{eq:originalFL}
m^2 \gtrsim g q \MPl^{\frac{d-2}{2}} H.
\end{align}
Here $m$ is the particle mass, $g$ is the gauge coupling of the $U(1)$ gauge theory, $q$ is the integer quantised charge of the particle, $\MPl$ is the reduced Planck mass and $H$ is the expansion rate (Hubble parameter) of the dS universe. This bound is obtained by applying the same low-energy physical principles that underly the Weak Gravity Conjecture (WGC) \cite{Arkani-Hamed:2006emk} to charged black holes in dS space. Namely, FL can be derived by demanding that large charged black holes (BHs) in dS space, called Nariai black holes, can evaporate while remaining subextremal. Despite potential caveats, this bottom-up derivation of FL is an appealing feature since it sidesteps the necessity for a quantum gravity construction of dS space. In addition, the FL conjecture~\eqref{eq:originalFL} withstands non-trivial checks in the context of string theory. For example, FL reproduces well-known consistency conditions when applied to D3-branes wrapping the A-cycle in a Klebanov-Strassler throat~\cite{Montero:2021otb}.

In the context of the WGC (see~\cite{Harlow:2022ich} for a review), one requires the existence of particles/branes with a mass-to-charge ratio that can be bounded above using properties of large extremal black holes~\cite{Arkani-Hamed:2006emk,Heidenreich:2015nta}:
\begin{align}
\label{eq:WGC}
 \alpha_{p,d} T_p^2 \leq e_{p,d}^2 q^2 M_d^{d-2}
\end{align}
where $\alpha_{p,d}$ is a known $\mathcal{O}(1)$ constant, $T_p$ is the tension of a $p$-brane, $e_{p,d}$ is the gauge coupling of the $U(1)$ $(p+1)$-form gauge field under which the brane is charged, $q$ is the integer brane charge and $M_d$ is the $d$-dimensional Planck mass. This bound can be motivated from low-energy considerations by requiring that large extremal black holes/branes can decay without becoming super-extremal. In this way, the WGC has an intimate connection to cosmic censorship~\cite{Penrose:1969pc}. See for instance~\cite{Horowitz:2016ezu,Crisford:2017zpi,Crisford:2017gsb} for a discussion of this connection in the context of asymptotically AdS spacetimes. Several other bottom-up arguments have been given for the validy of the WGC, for example, based on holography~\cite{Nakayama:2015hga,Harlow:2015lma,Benjamin:2016fhe,Montero:2016tif,Montero:2018fns}, thermodynamics~\cite{Banks:2006mm,Hod:2017uqc,Hebecker:2017uix,Cheung:2018cwt}, analyticity, unitarity and/or causality~\cite{Adams:2006sv,Hamada:2018dde}. In addition, it turns out that every string theory example indeed contains weak gravity particles/branes. Here, the top-down and bottom-up arguments come hand-in-hand and give us a lot of confidence in the validity of the WGC. 

With these two statements in mind, we have now come to the subject of this work. A superficial glance at equations~\eqref{eq:originalFL} and~\eqref{eq:WGC} reveals a major gap in the status of FL compared to the WGC and raises the natural question: is there a version of FL that also applies to branes in dS space? In this paper, we answer this question in the affirmative by studying the nucleation of branes around Nariai black holes. Following the FL logic, and using brane nucleation as an analog of Schwinger pair production, we argue for a series of bounds on the properties (i.e. tensions and certain couplings) of branes. Our derivations will mostly follow physical arguments and leave plenty of room for sharpening our results. In particular, we will not be able to fix any of the $\mathcal{O}(1)$ numbers that appear in the new FL bounds. 

Before stating the results of this paper, let us first clarify the set up we have in mind. We will study $p$-branes with a world-volume action that contains terms of the form:
\begin{align}
\label{eq:genericCoupling}
    I \supset T_p \int_{\rm WV} *1 + g_p \int_{\rm WV} A_1 \wedge \Omega_p
\end{align}
where $A_1$ is the bulk U(1) gauge field and $\Omega_p$ is a $p$-form constructed by taking wedge products of 1- or 2-form field strengths which may be intrinsic to the brane. For example, in the case of a 2-brane, we may take $\Omega_2 = F_2^{(B)}$ where $F^{(B)}$ is the field strength of a gauge field localized to the brane or $\Omega_p = F$ with $F = dA_1$. This coupling means that the brane can become charged under the bulk 1-form gauge field if $\Omega_p$ has a non-trivial integral along the brane spatial directions. This is different from the usual notion of charge, which does not rely on the presence of $\Omega_p$. Additionally, for the usual notion of charge, the dimensionality of the object is fixed by the form degree of the gauge field and a $p$-brane is always charged under a $(p+1)$-form gauge field. For this reason, a direct application of FL to branes is not possible because the Nariai black hole and the brane are charged under gauge fields of different degree. By using a coupling of the form shown in~\eqref{eq:genericCoupling}, we can consistently discuss screening of electric charges by higher dimensional branes.

The outcome of applying the FL argument to $p$-branes is a collection of bounds on the brane tension $T_p$ in terms of $g_p$, its coupling to electromagnetic fields and $H$, the Hubble rate of the de Sitter space. For branes with world-volume gauge fields and an action given in eq.~\eqref{eq:evenpAction} or eq.~\eqref{eq:oddpAction}, these bounds take the form:
\begin{align}
\label{eq:resultSummary1}
T_p^{\frac{1}{p+1} + \frac{4-p}{4}} \gtrsim g_{p} M_d^{\frac{d-2}{2}} H \quad \mathrm{or} \quad T_p^{\frac{4-p}{4}} \gtrsim g_{p} M_d^{\frac{d-2}{2}} \quad \text{for $p$ even and $p\neq 4$};\\
T_p^{\frac{1}{p+1} + \frac{3-p}{4}} \gtrsim g_{p} M_d^{\frac{d-2}{2}} H \quad \mathrm{or} \quad T_p^{\frac{3-p}{4}} \gtrsim g_{p} M_d^{\frac{d-2}{2}} \quad \text{for $p$ odd and $p\neq 3$}.
\end{align}
We will show that the D-branes of Type II string theory, which have couplings similar to eq.~\eqref{eq:evenpAction} or eq.~\eqref{eq:oddpAction}, satisfy these bounds. In addition, we study 2-branes without world-volume gauge fields. These couple to electromagnetic fields in the same way as axion domain walls and are described by the action in eq.~\eqref{eq:braneWV}. The tension of these branes is bounded by:
\begin{align}
\label{eq:resultSummary2}
    T_2 \gtrsim (g\Mpl H)^{3/2}
\end{align}
unless there is a light axion in the theory which couples to the same gauge fields by the usual $\theta F\wedge F$ interaction. 

The rest of this paper is organized as follows. In section~\ref{sec:FLReview}, we review the FL conjecture, emphasizing key aspects of the argument leading to the conjecture and noting a few caveats and potential loopholes. Readers already familiar with the FL bound literature may safely skip this section. After this review, we present the results outlined above in more detail. In section~\ref{sec:FLforBranes}, we study branes with world-volume gauge fields. We start by describing the nucleation of 2-branes and present the relevant instanton solution explicitly and the corresponding FL bound. We then distill this argument to its essential components, demonstrating how the correct bound for 2-branes can be derived using scaling arguments, which we generalise to higher-dimensional branes and spacetimes. We end this section with a discussion of additional brane couplings and why these do not alter our results. In section~\ref{sec:FLforBranesnoWVGF}, we propose a similar FL bound for branes without world-volume gauge fields and also present important loopholes in our argument. Section~\ref{sec:stringtheory} applies the bounds derived in section~\ref{sec:FLforBranes} to D-branes in string theory, showing that they satisfy these constraints in the weak coupling limit. Finally, in section~\ref{sec:conclusion}, we conclude and suggest promising directions for future exploration of the FL conjecture.

\section{Review of the Festina Lente conjecture}
\label{sec:FLReview}

We start by reviewing the FL conjecture in general dimension and take this opportunity to introduce the notation we will be using. One argues for the FL bound by studying the evaporation of large charged black holes in dS space. These are solutions to Einstein-Maxwell theory in dS space called Nariai black holes and they can shed their mass by the usual Hawking process, emitting neutral particles such as gravitons or photons. In addition, they can shed their charge by a Schwinger process which screens the electric field. In the presence of light charged matter, Schwinger pair production can become very fast and leads to the electric field being screened on time scales much shorter than the Hawking radiation time scale. In particular, the Schwinger rate is set by the magnitude of the electric field which is comparable to $M_d^{\frac{d}{2}-1} H$ where $M_d$ is the $d$-dimensional Planck constant and $H$ is the Hubble rate, as we will see below. On the other hand, the Hawking emission rate is set by the temperature of the black hole which is of order $H$. Using this observation,~\cite{Montero:2019ekk} argued that the Nariai black hole spacetime becomes super-extremal in the presence of light charged particles. We will go through the argument in more detail below and also mention some potential caveats\footnote{For additional new work on this conjecture in the context of the early universe, or regarding confinement, see \cite{Venken:2023hfa, Mishra:2022fic,Lee:2021cor}, in which the \textit{inapplicability} of FL in certain regimes plays a key role.}. 

First, we present the black hole solutions of Einstein-Maxwell theory in dS space in arbitrary dimensions. The action including a cosmological constant $\Lambda$ is:
\begin{equation}
    \label{eq:EinsteinMaxwell}
    I=\int d^d x \sqrt{-g}\left(\frac{1}{16\pi G}\left(R-2\Lambda\right)-\frac{1}{4}F_{\mu\nu}F^{\mu\nu}  \right),
\end{equation}
where $F_{\mu\nu} = \pd_\mu A_\nu - \pd_\nu A_\mu$ is the electromagnetic field strength, $R$ is the Ricci scalar and $\Lambda$ is the cosmological constant. The $d$-dimensional Planck mass is related to the gravitational constant by $(8\pi G)^{-1} = M_d^{d-2}$ and the dS length $\ell^2 = \frac{(d-1)(d-2)}{2\Lambda}$.

This theory has spherically symmetric solutions of the following form:
\begin{equation}
    \label{eq:RNdSSolution}
    ds^2=-h(r)dt^2+h(r)^{-1}dr^2 +r^2 d\Omega^2_{d-2},
\end{equation}
where $d\Omega^2_{d-2}$ is the area element of a unit $(d-2)$-sphere. The blackening factor is given by:
\begin{equation}
    h(r)=1-\frac{16\pi G M}{(d-2)S_{d-2}}\frac{1}{ r^{d-3}}+\frac{8\pi G Q^2}{(d-3)(d-2)}\frac{1}{r^{2(d-3)}}-\frac{2 \Lambda}{(d-1)(d-2)}r^2,
\end{equation}
where $S_{d-2}$ is the total area of a unit $(d-2)$-sphere. These solutions also feature a non-zero electromagnetic gauge potential given by:
\begin{equation}
    A=-\frac{Q}{(d-3)S_{d-2}r^{d-3}}dt.
\end{equation}
This line element and gauge field configuration describe a black hole with mass parameter $M$ and electric charge parameter $Q$ in an asymptotically dS spacetime.
\begin{figure}[h]
\centering\includegraphics[width=0.65\textwidth]{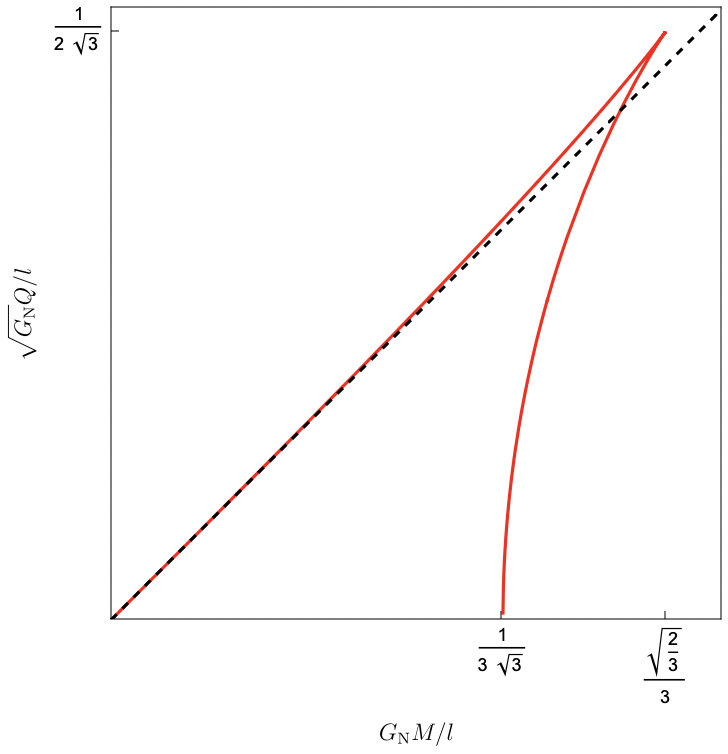}
\caption{The \textit{shark fin} diagram depicting in red the lines of extremal black holes solutions, where at least two horizons meet ($d=4$ here for concreteness). The ultracold point is at the tip of the shark fin. The diagonal black dashed line represents points satisfying $\sqrt{G_{\mathrm{N}}}M=Q$.}
\label{fig:ParametricPlotNariai}
\end{figure}

Let us proceed by describing important features of these black hole solutions. The space of solutions is shown in Fig.~\ref{fig:ParametricPlotNariai} and each point represents a spacetime that may have up to three horizons: an inner and an outer black hole horizon, and the cosmological horizon. As usual, values of the parameters $M$ and $Q$ for which at least two horizons coincide represent `extremal' solutions. In appendix \ref{appA}, we describe these extremal solutions that represent the `shark fin' boundary between the sub-extremal and super-extremal regions in Fig. \ref{fig:ParametricPlotNariai}. The upper red line connected to the origin, or the \textit{upper branch}, represents the mass and charge parameters for which the inner and outer black hole horizons coincide (the black hole near-horizon becomes $\mathrm{AdS_2} \times S^{d-2}$, see appendix \ref{appA}) while the cosmological horizon stays far away (except at the tip of the shark fin known as the \textit{ultracold point}, where the near-horizon geometry becomes $\mathrm{Mink}_2\times S^{d-2}$). These are the extremal Reissner-Nordstr\"{o}m black holes in a de Sitter universe. The black dashed line of Fig.~\ref{fig:ParametricPlotNariai} represents extremal Reissner-Nordstr\"{o}m solutions of an asymptotically flat background; this line would represent extremal black holes if the action had vanishing cosmological constant. Indeed, in an asymptotically flat universe, there is no limit to how large an extremal Reissner-Nordstr\"{o}m black holes could be. But in a dS universe, if we  were to make an extremal black hole larger and larger, eventually the black hole would become large enough to reach the cosmological horizon. This is the ultracold point, which is the tip of the shark fin diagram. Additionally, the red right branch represents the case in which the black hole outer horizon coincides with the cosmological horizon (in the coordinates of eq.~\eqref{eq:RNdSSolution}) but the inner horizon does not\footnote{At the ultracold point all three horizons coincide, and on the $Q = 0$ axis the inner black hole horizon is at $r=0$.}. This is the \textit{Nariai branch}. These Nariai black holes (which have a $\mathrm{dS}_2\times S^{d-2}$ geometry) are the main focus of the Festina Lente bound and will play a similarly important role in this paper. We outline in appendix~\ref{appA} the origin of the numerical values labelling special points on Fig.~\ref{fig:ParametricPlotNariai}. 

While points on the shark fin represent extremal solutions, points in the interior represent sub-extremal solutions, and points on the exterior represent super-extremal solutions. In particular, we note that if one starts with a solution on the Nariai branch and allows this solution to discharge too quickly, then the spacetime may leave the shark fin region. It is precisely this feature that allows one to argue for the Festina Lente conjecture \cite{Montero:2019ekk, Montero:2021otb} as we now review.
\begin{figure}[h]
\centering\includegraphics[width=1.0\textwidth]{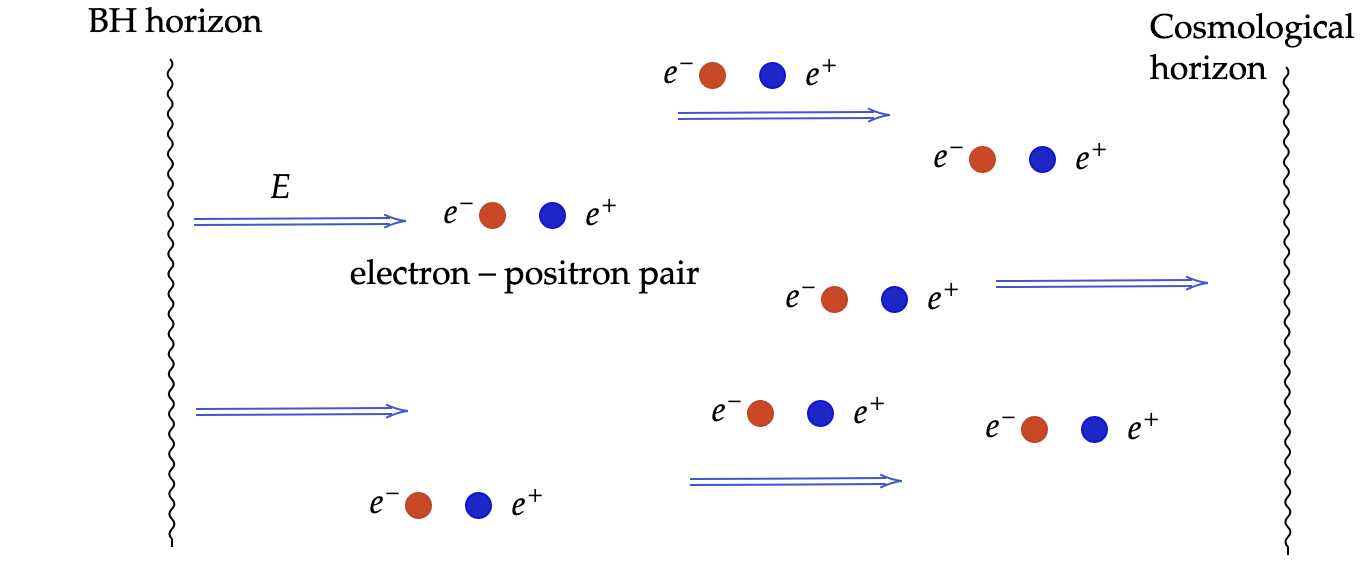}
\caption{Schematic diagram showing pair nucleation - such as electrons and positrons - in a background electric field $E$, on the Nariai branch.}
\label{fig:fl1}
\end{figure}
Suppose that, in addition to the theory described by eq.~\eqref{eq:EinsteinMaxwell}, we also have charged matter with integer charge $q$ and mass $m$. In the presence of charged matter, the electric field of a Nariai black hole is unstable to Schwinger pair production (see Fig.~\ref{fig:fl1}) which discharges the black hole. For instance, one may measure the black hole charge using the electric flux through a sphere surrounding the black hole (i.e. Gauss' Law). As the electric field is screened, the electric flux and black hole charge decrease. When this process is slow, the pair production rate\footnote{Here we are quoting the Schwinger rate in flat space. As we mentioned above, the Nariai spacetime has a $dS_2 \times S^2$ geometry. Effects of this geometry were considered in \cite{Frob:2014zka,Montero:2019ekk,Montero:2021otb} but these do not modify the exponent which is the essential part of the FL argument. More physically, one can consider that the pair production process happens locally (on distances set by $m/gq\Mpl H$) and is thus insensitive to the small spacetime curvature (which is set by $H^{-1}$).} per volume is given by\footnote{For additional work on the Schwinger effect in compact space and/or finite temperature, see also~\cite{brown2018schwinger,Frob:2014zka,Gould:2017fve,Gould:2018efv,Draper:2018lyw,Samantray:2020uzp,Qiu:2020gbl}.} \cite{Schwinger:1951nm, Heisenberg:1936nmg, brown2018schwinger, Ai:2020vhx, Brown:1988kg}:
\begin{align}
\label{eq:SchwingerRate}
\Gamma/V \sim (gqE)^{d/2} e^{-\pi m^2/gqE}.
\end{align}
where $E \sim M_d^{\frac{d-2}{2}} H$ is the electric field of the Nariai black hole and $g$ is the gauge coupling of the U(1) theory. This expression also quantifies the meaning of `slow' in the previous statement. As usual in the case of non-perturbative processes, rates computed using instanton methods are trustable only in the region where the instanton action is $\gg 1$. When this condition is satisfied the process is slow. From eq.~\eqref{eq:SchwingerRate}, we see that the condition for slow discharge is:
\begin{align}
\label{eq:FLbound}
m^2 \gtrsim g q M_d^{\frac{d-2}{2}} H
\end{align}
where we substituted the value of the electric field of a Nariai black hole. 

In the opposite limit, i.e. when\footnote{For even lighter fields, satisfying $m \ll H$,~\cite{Frob:2014zka} showed that Schwinger pair production in 1+1 dimensional de Sitter space leads to infrared hyperconductivity where a small electric field can drive a large current.} $m^2 \ll g q M_d^{\frac{d-2}{2}} H$, one cannot trust the rate given by eq.~\eqref{eq:SchwingerRate}. Nevertheless, one expects the Schwinger pair production rate to be fast and the electric field to rapidly discharge into particle-antiparticle pairs that annihilate to produce photons. Moreover, as $m \rightarrow 0$, one expects the rate to be set only by the electric field. This is suggested by eq.~\eqref{eq:SchwingerRate} where the prefactor is set only by the electric field. Of course, as mentioned earlier, the semi-classical approximation giving eq.~\eqref{eq:SchwingerRate} is only valid for large $m$ (more precisely, a large exponent) and should not be trusted in the limit $m\rightarrow 0$. Nonetheless, in the limit $m \rightarrow 0$, the only remaining scale in the problem is the electric field and dimensional analysis leads to the same conclusion, i.e. the rate should be set by the electric field only\footnote{In higher dimensions, the gauge coupling $g$ also carries units and may also appear in the prefactor. However, the only relevant combination is the product $gE$ which can be seen explicitly in the analysis of the following sections.}. 

This argument implies that the discharge rate is set by $g M_d^{\frac{d-2}{2}} H$ rather than the characteristic time scale of the spacetime (including the Hawking radiation time scale) which is the Hubble rate $H$. Since this discharge process is much faster than the Hawking evaporation process, \cite{Montero:2019ekk} argue that the evolution is captured by replacing the electric field by a gas of photons\footnote{As discussed in~\cite{Montero:2019ekk}, this choice does not affect the final result and one may replace the electric field with a thermal bath of particles with any reasonable equation of state (not just that of photons).} with the same energy density. This is because the particles and anti-particles produced by the electric field can locally annihilate into photons. Although no calculation has been performed to illustrate this annihilation process, the conclusion is plausible given the following thought experiment. Suppose we start with two separate regions, called $A$ and $B$ say, that have a non-zero electric field. Both of these regions can produce particle/anti-particle pairs that eventually leave their respective region. Once the particles leave their respective region they may interact with a detector or particles from the other region. In particular, if a particle from region $A$ meets an anti-particle from region $B$, the two may annihilate into photons outside either region. Suppose now that we shrink the gap between $A$ and $B$ (or fill that gap with an electric field), then one would expect this production and annihilation process to still happen. This is basically the assumption made in \cite{Montero:2019ekk} and we will continue to assume\footnote{In fact, the thought experiment outlined above may be one way to tackle the question of pair production and annihilation more precisely. For instance, one may calculate the Schwinger current for particles and anti-particles from regions $A$ and $B$ and use this result to find the annihilation rate between a current of particles and an opposing current of anti-particles. We leave this for future work.} it here.

With the above assumption, one can track the evolution of the Nariai spacetime by starting with a $dS_2 \times S^2$ spacetime filled with a gas of photons. The initial metric is:
\begin{align}
ds^2 = e^{-\phi(t)}(-dt^2 + a(t)^2 dx^2) + e^{2\phi (t)}d\Omega^2
\end{align}
and the Einstein equations give the time evolution of $a(t)$ and $\phi(t)$ (see Appendix~\ref{sec:appendixDyonicBHs} for the equations). In addition, current conservation gives the evolution of the photon energy density. Solving this system of equations, it is easy to check that $\phi \rightarrow 0$ within a Hubble time. This means that the sphere collapses indicating that `the whole spacetime has fallen into the black hole'. This solution is similar to the Big Crunch solution one would get on the $Q=0$ axis for black hole masses higher than the Nariai limit and for observers inside this would-be black hole. Since we consider these solutions super-extremal (i.e. they have a mass higher than what one would consider allowed in dS space), then one should demand that time evolution of a reasonable spacetime (i.e. the Nariai spacetime) does not end up with this super-extremal solution. The only way to ensure this is to have all charged matter satisfy eq.~\eqref{eq:FLbound}. This is the FL bound.

Before we proceed, let us comment on some potential caveats. First, the Big Crunch spacetime that one gets by evolving the Einstein equations has a space-like singularity and it is unclear whether this is pathological behaviour. One reason to exclude this behaviour is to rely on the cosmic censorship conjecture~\cite{Penrose:1969pc} to exclude all spacetimes outside the shark fin region in Fig.~\ref{fig:ParametricPlotNariai}. For most of these spacetimes, the singularity is time-like and naked which is deemed pathological by cosmic censorship. On the $Q=0$ axis, however, the singularity is space-like and it is not obvious if cosmic censorship should extend to these cases. That said, one can imagine that the screening process does not go to completion but leaves the black hole with a small charge. Indeed, the process will become slow when the electric field drops to a value such that:
\begin{align}
E \lesssim \frac{m^2}{g q}
\end{align}
and the Schwinger process is again exponentially suppressed. If the black hole is left with a small charge, and the spacetime is indeed super-extremal after this rapid discharge, then there may be a naked time-like singularity which contradicts cosmic censorship. This claim should be substantiated by further calculations and we believe more work is needed to understand the relation between cosmic censorship and the FL bound.

Additionally, cosmic censorship is now understood as a statement about generic initial conditions~\cite{Geroch:1979uc}. For instance, fine-tuned field configurations can produce naked singularities (see for example~\cite{Choptuik:1992jv,Christodoulou:1994hg,Gundlach:2007gc,Bizon:2011gg}). It could be the case that tuned initial conditions leading to naked singularities are not in contradiction with cosmic censorship. In this spirit, one may claim that starting with a Nariai black hole is a tuned initial condition. However, one may always start with a black hole that has a charge slightly higher than that of a Nariai black hole (this is always possible on the Nariai branch). In the presence of light charge particles these spacetimes can be quickly driven to Nariai spacetimes (in a process analogous to what happens to extremal black holes which are driven towards extremality in the absence of light charged matter \cite{Hiscock:1990ex}). If subsequent evolution of these Nariai black holes is described as in~\cite{Montero:2019ekk}, it may be possible to extend FL to apply to an open set of initial conditions. This would evade any necessary fine-tuning of initial conditions and cosmic censorship should then be valid. In this context, the view advocated in~\cite{Lin:2024jug} could play an important role. Namely, it is possible that the light charged particles in the spectrum modify the black hole solutions in the region near the Nariai line in Fig.~\ref{fig:ParametricPlotNariai} so that these black holes always need to be considered with a surrounding cloud of charged particles. This cloud could alter the evolution of black holes as they approach the Nariai line and prevent them from exiting the shark fin region (see also \cite{Ong:2019glf}). While still on the topic of cosmic censorship, we briefly comment that string theory may resolve apparently naked singularities in general relativity in unexpected ways. There are examples~\cite{Horowitz:1995tm,Breckenridge:1996is,Gimon:2004if} where spacetimes appear to have naked singularities at large distance but these are cloaked by stringy ingredients so no horizons might be necessary to hide such singularities. In analogy with these examples, one may speculate that string theory provides a resolution of the FL pathologies that does not require imposing the bound in eq.~\eqref{eq:originalFL}. 

Another potential caveat with the FL argument has been raised in connection with spacetime back-reaction effects. This is described in~\cite{Aalsma:2023mkz} by studying black hole decay using the tunneling formalism~\cite{Parikh:1999mf}. By focusing on a single decay channel (s-wave single-particle production),~\cite{Aalsma:2023mkz} finds no evidence that Nariai black holes exit the shark fin if back-reaction effects are taken into account even when this decay mode is fast. This analysis has not yet been extended to the multi-particle case, including particle--anti-particle annihilation effects, which is the relevant regime for the FL bound. As such, the status of back-reaction effects remains unclear although the conclusions of~\cite{Aalsma:2023mkz} suggest that these could be important in understanding the FL bound. Ultimately, whether the black hole exits the shark fin region is still an open question that we believe warrants further exploration. In the absence of a complete understanding of this process, we may treat this behaviour as an assumption in our analysis.

With these and other\footnote{For example, the interpretation of the $\Mpl \rightarrow \infty$ limit remains ambigious in the context of FL since the bound does not trivialise in that limit. We do not have more to add to this beyond what has already been said in the literature \cite{Montero:2019ekk,Montero:2021otb}.} caveats in mind, it is evident that a more precise calculation of the FL effect is highly valuable. Although we do not pursue this challenging task here, we instead explore the FL bound by extending its logic to brane sources and validating the resulting bounds on known examples. The success of these checks increases our confidence in the FL argument. This is the approach taken in this paper, and the reader should note that our statements carry the same caveats as the original FL bound.

\section{Branes with world-volume gauge fields}
\label{sec:FLforBranes}

The discussion above highlights the important aspects of the FL argument and how one may extend it to theories with a more general content (i.e. strings, branes, etc.). Specifically, we saw that the screening of the electric field and its efficient conversion to a bath of particles is the reason behind the FL conjecture. Based on this observation, it follows that the bound can be extended to any process that rapidly screens the electric field and efficiently converts its energy density to a non-coherent bath of particles. In this section, we will consider the nucleation of branes in the charged Nariai spacetime which provides a mechanism to screen the electric field analogous to what happens with Schwinger pair production. For certain values of the brane tension and couplings, the result of this nucleation is a fast reduction in the black hole charge. Once the electric field is screened and converted to an incoherent energy density, the $dS_2 \times S^2$ spacetime crunches and we interpret this Big Crunch as a spacetime outside the extremality region in Fig.~\ref{fig:ParametricPlotNariai}. Using the logic of the FL conjecture, we require that this process is suppressed which gives bounds on the tension of these branes in terms of their couplings to the 1-form gauge fields\footnote{As we will shortly see, the branes we consider are not charged in the usual sense but couple via a Wess-Zumino term to a 1-form gauge field. This coupling plays the role of the `charge' in our derivations.}. 

Before proceeding, we comment on the level of detail provided in the following analysis. If we were to study the evolution of the $dS_2 \times S^2$ solution in full detail, we would have to calculate the bubble nucleation, expansion and collision. For the collisions, we would have to describe the conversion of the initial electromagnetic energy density to particles, string, branes, etc. In practice, studying this creation and annihilation process is prohibitively difficult and has not been done even for the simpler Schwinger pair-production. Since this process is more complicated for branes, we will not explore this direction in what follows. Instead, we will assume that this creation and annihilation process occurs and converts an $\mathcal{O}(1)$ fraction of the energy in the electromagnetic field to non-coherent sources (i.e. particles, strings, etc.). This is an important assumption and if it proves to be inaccurate, it can affect the conclusion of our analysis. On the other hand, while we are also agnostic to the end products of these brane collisions, we do not expect this detail to affect our results significantly. In particular, different end products will lead to different equations of state for the non-coherent bath and this was shown to not impact the conclusion of the FL argument in~\cite{Montero:2019ekk}. As mentioned previously, instead of tackling these and other potential issues (see section~\ref{sec:FLReview}), we proceed with an exploration of the consistency of the FL argument when applied to extended objects and compared to the extended objects we have in string theory. Remarkably, FL passes this check.

This section is organised as follows. First, we will consider 2-branes with a Wess-Zumino world-volume interaction. This interaction provides a coupling between a bulk 1-form gauge field to a gauge field intrinsic to the brane. We will show that one can find an explicit Euclidean solution describing the nucleation of these branes with a non-trival world-volume gauge field profile. This non-zero profile means that the nucleated brane is charged under the bulk gauge field and can screen the electric field of the Nariai black hole (which is itself charged under the bulk 1-form gauge field). Requiring that this process is slow, in the sense defined in the previous section, gives a new bound on the tension of these branes. Following the 2-brane discussion in 4 dimensions, we generalize our result to $p$-branes in $d$-dimensions.

\subsection{Nucleation of 2-branes}
\label{sec:2branes}

An essential ingredient in the Festina Lente argument is the rate for Schwinger pair production eq.~\eqref{eq:SchwingerRate}. This rate may be derived from a worldline formalism that has the advantage of being easily generalized to higher-dimensional objects \cite{Brown:1988kg,Feng:2000if,brown2018schwinger,Ai:2020vhx}. In this section, we will carry out this derivation for 2-branes explicitly. In particular, we will calculate the nucleation rate of 2-branes in the presence of a background field which is a constant uniform electric field in the $z$ direction ($\mathbf{E}=E\mathbf{\hat{z}}$) -- just as in the typical Schwinger effect. For the 2-brane world volume (WV) action:
\begin{align}\label{d2action}
    I=&\int_{\mathrm{WV}} d^3\xi \sqrt{-h} \Bigg(-\frac{1}{4}F^{(\mathrm{B})}_{ab}F^{(\mathrm{B})ab}\Bigg)-T\int_{\mathrm{WV}} d^3\xi \sqrt{-h}+g\int_{\mathrm{WV}}\sqrt{-h}\epsilon^{abc}A_a F^{(\mathrm{B})}_{bc} \nonumber\\
    &+\int_{\mathrm{Bulk}} d^4x\Bigg(-\frac{1}{4}F_{\mu\nu}F^{\mu\nu}\Bigg).
\end{align}
In Euclidean signature, this action corresponds to:
\begin{align}\label{d2actionEuclidean}
    I_{\mathrm{E}}=&\int_{\mathrm{WV}} d^3\xi \sqrt{h} \Bigg( \frac{1}{4}F^{(\mathrm{B})}_{ab}F^{(\mathrm{B})ab}\Bigg)+T\int_{\mathrm{WV}} d^3\xi \sqrt{h}-ig\int_{\mathrm{WV}}\sqrt{h}\epsilon^{abc}A_a F^{(\mathrm{B})}_{bc}\nonumber\\
    &+\int_{\mathrm{Bulk}} d^4x\Bigg(\frac{1}{4}F_{\mu\nu}F^{\mu\nu}\Bigg).
\end{align}
Here $A_\mu$, with $F=dA$, is the bulk electromagnetic vector potential, and $A^{(\mathrm{B})}_{a}$, with $F^{(\mathrm{B})}=dA^{(\mathrm{B})}$ is a $U(1)$ gauge potential living on the brane. The constant bulk electric field configuration implies that $F_{t_\mathrm{E} z}=iE$, and if we gauge fix $A_{\mu}$ such that only $A_{t_{\mathrm{E}}}$ is non-vanishing, then it follows that a solution of the vector potential is $A_\mu=iEr\cos{\sigma}\delta^{t_\mathrm{E}}_\mu$. Treating this term as a background field, we now now have to solve for the gauge field $A^{(\mathrm{B})}$ corresponding on the surface of a nucleated spherical bubble.

The equation of motion for the brane gauge field is 
\begin{equation}\label{eomAB}
    \partial_a\Big(\sqrt{h}F^{(\mathrm{B})ab}-2ig\sqrt{h} \epsilon^{cab}A_c\Big)=0.
\end{equation}
Physically, this equation states that the background bulk electric field induces charges on the brane, since we can read it as a modified Gauss law. To properly fix the conventions, let us now establish a coordinate system compatible with the standard rotationally symmetric ansatz. Following \cite{Coleman:1977py, Coleman:1985rnk, Coleman:1977th,Callan:1977pt} we assume that there is a solution satisfying 
\begin{equation}
r^2+t_{\mathrm{E}}^2=r_*^2,
\end{equation}
in Euclidean signature\footnote{In real time, $r^2-t^2=r_*^2$, corresponding to hyperbolic motion.}, with $x^2+y^2+z^2=r^2$ i.e. the $O(4)$ of Coleman \cite{Coleman:1977py,Callan:1977pt,Coleman:1977th,Coleman:1980aw,Coleman:1985rnk}.
We take the bulk coordinate system to consist of $\{t_\mathrm{E},r,\sigma,\phi\}$; in the bubble interior, we take the coordinate system to be $\{\tau_\mathrm{E}, R,\sigma,\phi\}$, and the corresponding line elements are respectively:
$$ ds_{\mathrm{out}}^2=dt_{\mathrm{E}}^2+dr^2+r^2d\sigma^2+r^2 \sin^2{\sigma}d\phi^2,$$
$$ ds_{\mathrm{in}}^2=d\tau_{\mathrm{E}}^2+dR^2+R^2d\sigma^2+R^2 \sin^2{\sigma}d\phi^2,$$
the latter of which extends up to the bubble.
On the bubble, these two line elements must match\footnote{This matching procedure is akin to Oppenheimer-Snyder collapse of a thin shell (in real time of course), except the situation here is simpler since we take the exterior metric to be flat rather than Schwarzschild.}, forcing $r=R$ and $dR=0$, as well as
\begin{equation}
    \Big(\frac{dr}{d\tau_\mathrm{E}}\Big)^2+\Big(\frac{dt_\mathrm{E}}{d\tau_\mathrm{E}}\Big)^2=1.
\end{equation}
A solution compatible with our aforementioned ansatz is 
\begin{equation}\label{randt}
    r(\tau_{\mathrm{E}})=r_* \cos{(\tau_{\mathrm{E}}/r_*)}, \ \mathrm{and}\quad t_{\mathrm{E}}(\tau_{\mathrm{E}})=r_* \sin{(\tau_{\mathrm{E}}/r_*)}.
\end{equation}
It is a choice to have set our clocks such that $t=0$ corresponds to $\tau=0$.\footnote{Note in real time, $r(\tau)=r_* \cosh{(\tau/r_*)}, \ \mathrm{and}\quad t(\tau)=r_* \sinh{(\tau/r_*)}.$ This is the trajectory of a uniformly accelerating observer, which is once again expected.}
The picture in Euclidean signature is clear: we have an $O(4)$ instanton appearing at $\tau_{\mathrm{E}}=0$, which we continue by Wick rotation to Lorentzian time and match at $\tau=0$. The nucleated bubble that is born with a critical radius $r_*$ - which we will compute momentarily - then expands in a hyperbolic trajectory which asymptotically approaches a future light cone. 

Using the parametrization of eq.~\eqref{randt}, the projection relating the bulk (flat) coordinates to the coordinates on the brane now have the following components:
\begin{equation}
    e^{t_{\mathrm{E}}}_{\tau_{\mathrm{E}}}=\cos{(\tau_{\mathrm{E}}/r_*)},\quad e^r_{\tau_{\mathrm{E}}}=\sin{(\tau_{\mathrm{E}}/r_*)},\quad e^\sigma_\sigma=1,\quad e^\phi_\phi=1,
\end{equation}
with all others vanishing. By projecting the bulk metric, we find that the metric on the brane with intrinsic coordinates $\{\tau_{\mathrm{E}}, \sigma,\phi\}$ is
\begin{equation}
    h_{ab}=
    \begin{pmatrix}
1 & 0 & 0\\
0 & r(\tau_{\mathrm{E}})^2 & 0\\
0 & 0 & r(\tau_{\mathrm{E}})^2\sin^2{\sigma}
\end{pmatrix}.
\end{equation}
We now possess all of the ingredients necessary to solve eq.~\eqref{eomAB}. Note that by azimuthal symmetry, we should be able to remove dependence on the equatorial angle $\phi$ in $A^{(\mathrm{B})}_a$, i.e. $A^{(\mathrm{B})}_a=A^{(\mathrm{B})}_a(\tau_{\mathrm{E}},\sigma)$. Therefore, we are free to gauge fix $A^{(\mathrm{B})}_\sigma=0$; with this symmetry and the given ansatz, we would not have been able to gauge away $A^{(\mathrm{B})}_\phi$. Recalling that $A_{\tau_{\mathrm{E}}}=A_{t_{\mathrm{E}}} e^{t_{\mathrm{E}}}_{\tau_{\mathrm{E}}}$ (with other components vanishing), due to the anti-symmetry of the interaction term, $A^{(\mathrm{B})}_{\tau_{\mathrm{E}}}$ does not appear; a solution satisfying $\partial_\sigma A_{\tau_{\mathrm{E}}}\sim \mathrm{constant}/\sin{\sigma}$ is possible but such a term gives the brane kinetic piece an infinite Euclidean action, and so only a vanishing constant prefactor is allowed. 

Now, we turn to the problem of solving for $A^{(\mathrm{B})}_\phi$, which we have argued is the only non-vanishing component of $A^{\mathrm{(B)}}_a$. Setting the free index in eq.~\eqref{eomAB} to $\phi$ gives the following second order partial differential equation: 
\begin{align}\label{eomAphi}
    \partial_\sigma\Bigg( \frac{1}{r_*^2 \cos^2{(\tau_{\mathrm{E}}/r_*)} \sin\sigma} \partial_\sigma A^{(\mathrm{B})}_{\phi} -2gEr_* \cos^2{(\tau_{\mathrm{E}}/r_*)} \cos{\sigma} \Bigg)=-\frac{1}{\sin \sigma}\partial_{\tau_{\mathrm{E}}}^2 A^{(\mathrm{B})}_{\phi}.
\end{align}
It is possible to find a solution to this equation:
\begin{equation}\label{aphisol}
A^{(\mathrm{B})}_{\phi}=K\cos^2{(\tau_{\mathrm{E}}/r_*)}\sin^2{\sigma}, 
\end{equation}
with an unkown coeffiecient $K$. We now use this informed guess in eq.~\eqref{eomAphi}. Hearteningly, this ansatz is indeed a solution.\footnote{Originally, we solved this by analogy with the axion domain wall nucleation process studied in~\cite{Hassan:2024nbl}. There, it was shown that parallel electric and magnetic fields are unstable to the nucleation of axion domain walls and an instanton was found that describes this process. This analogy is useful because thin axion domain walls have a world volume action with a Chern-Simons term that is similar to the Wess-Zumino term we have in eq.~\eqref{d2actionEuclidean}, up to the replacement $F^{\rm (B)} \rightarrow F$. Since we know the solution for $F$, which is a constant electric and magnetic field in the weak coupling limit, we can choose a similar solution for $F^{\rm (B)}$ that will solve the above equation. Since the background $A_\mu$ already contains the electric piece, we choose $F^{\rm (B)}$ to mimic the magnetic field in the $\hat{z}$ direction. That is precisely the form of $A^{(\mathrm{B})}_{\phi}$ we have here. See \cite{Hassan:2024nbl} for details.} In fact, this step fully fixes the numerical prefactor $K=-gEr_*^3/2$ as well. We emphasise once more for clarity that there is no background magnetic field, just an electric field; the magnetic field is only used as a trick for solving for the vector potential $A^{\rm (B)}$. 
\begin{figure}[h]
\centering\includegraphics[width=0.6\textwidth]{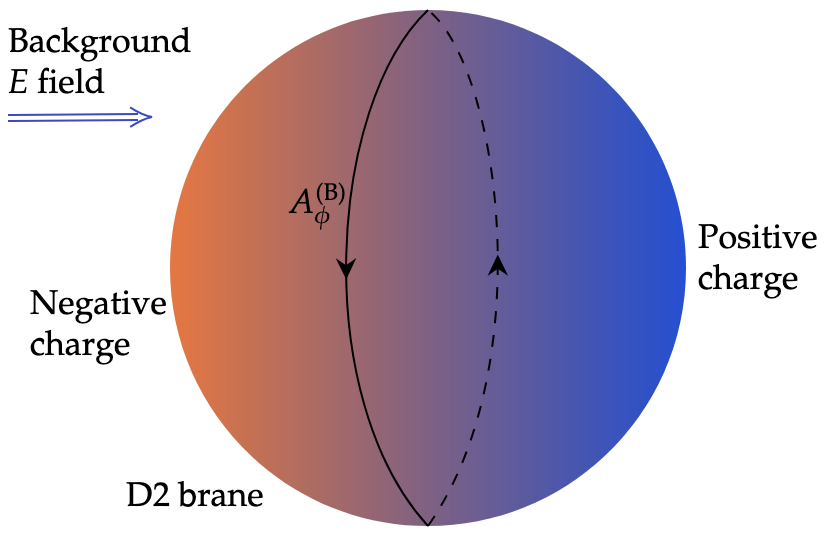}
\caption{Schematic diagram representing brane nucleation. The oppositely colored hemispheres indicate that they are oppositely charged. There is a non-vanishing gauge field $A_\phi^{(\mathrm{B})}$ on the brane.}
\label{fig:fl4}
\end{figure}

In summary, we have shown a solution to eq.~\eqref{eomAB} corresponding to a spherical brane bubble nucleation process. We may also like to ask how the background electric field back-reacts to this nucleation process. As we have briefly mentioned before, the Wess-Zumino interaction assigns electric charges to the brane, but as the bubble expands in real time, there will be currents due to its expanding nature, and hence magnetic fields as well. An analogous step occurs in the Schwinger effect when the charges particles accelerate. As such, let us then only focus on the induced charges at nucleation, i.e. at $\tau_{\mathrm{E}}=\tau=t=0$.

Looking once again at the action eq.~\eqref{d2action}, we can treat the coefficient of $A_a$ as a source for the electromagnetic gauge field, which is to say that we can treat $g\epsilon^{abc}F^{(\mathrm{B})}_{bc}$ as an electric source current $j^a$. We can then solve the (sourced) wave equations\footnote{It is helpful to now return to real time.} in the bulk upon converting $j^a \rightarrow j^{\alpha}$ with the projection tensors $e^\alpha_a$, thereby extracting the back-reaction on the electric field; this effect will be higher order in $g$, of course. We find that at $t=0$, the only non-vanishing component of the current is 
\begin{equation}
    j^t = 2g^2 E r_* \cos\sigma,
\end{equation}
which we identify as the elctric charge density, $\rho_{\mathrm{E}}$, localized by a radial delta function on the brane. In essence, the brane acquires the charge distribution resembling a conducting thin shell in a uniform background electric field in the $\mathbf{\hat{z}}$ direction. The northern and southern hemispheres are oppositely charged. At the moment of birth, the electric field gets screened uniformly in the interior, and the electric potential is proportional to a first order Legendre polynomial, with radial scaling behavior of the form $\sim r$ and $\sim r^{-3}$ in the bubble interior and exterior respectively. It is this screened electric field that provides the energy for the brane, and the gauge field configuration living on the brane. Upon nucleation, the background electric field pulls apart these two oppositely charged hemispheres, leading the bubble to expand.

We are left with one final task: we must still calculate $r_*$. Towards this end, we put the solutions we have obtained so far in the Euclidean action eq.~\eqref{d2actionEuclidean} and minimize it with respect to $r_*$, which in turn yields the critical radius of the bubble. We find that 
\begin{equation}\label{r_*}
    r_*\sim \frac{T^{1/2}}{|gE|}.
\end{equation}
We may now obtain the nucleation rate:
\begin{equation}\label{d2rate}
    \Gamma/V \sim  \exp{(-I_{\mathrm{B}})}, \quad \mathrm{with}\quad I_{\mathrm{B}}\sim \frac{T^{5/2}}{(gE)^3} .
\end{equation}
Here we compute the exponent but not the determinant prefactor which we assume is non-vanishing. Moreover, we assume throughout that there is a single negative eigenvalue for this fluctuation determinant so giving the usual interpretation of a physical decay rate. In dS space, the rate of processes that screen the electric field, such as the nucleation of the 2-branes considered here, cannot be too large (in the sense discussed in section~\ref{sec:FLReview}). This is to avoid pathological black hole decays if one starts with a charged Nariai black hole. For the rate we computed in eq.~\eqref{d2rate}, this means that the exponential factor has to be small when considered in a charged Nariai black hole background. For a generic charged Nariai black hole, the electric field is $E \sim \Mpl H$. Requiring exponential suppression of the nucleation rate means that:
\begin{align}
\label{eq:exponentialSuppression}
T \gtrsim (g \Mpl H)^{6/5}.
\end{align}
\begin{figure}[h]
\centering\includegraphics[width=0.9\textwidth]{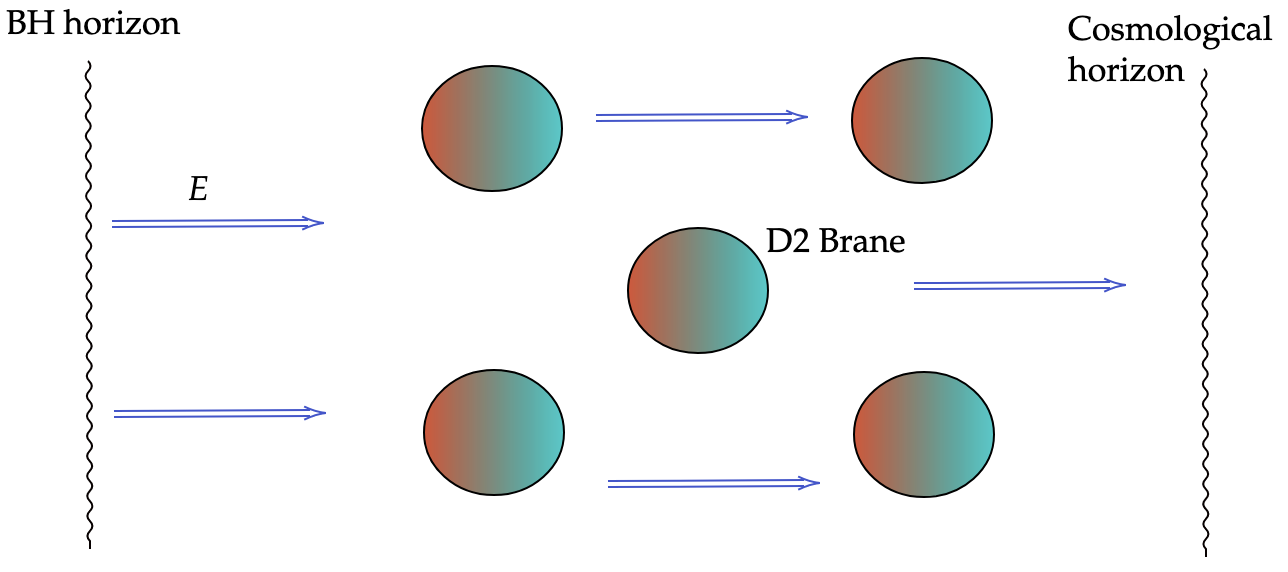}
\caption{Schematic diagram showing brane nucleation on the Nariai branch.}
\label{fig:fl3}
\end{figure}
Either this has to be true or the radius computed in eq.~\eqref{r_*} has to be larger than the size of the $S^2$ so that the flat space limit we are considering no longer makes sense, and the bubble is of similar size to the curvature scale; we may view this as an obstruction to nucleation since a bubble that is too large will not fit. On the Nariai solutions, the size of the $S^2$ is comparable to $H^{-1}$ and the inequality $r_* \gtrsim H^{-1}$ translates to another lower bound on the brane tension. We note that this upper bound is independent of $H$ since these factors cancel on either side. Since either of these two conditions is sufficient to avoid the FL contradiction, we get the following inequality:
\begin{align}
\label{eq:FL2brane}
T\gtrsim \min \left[(g \Mpl H)^{6/5}, (g \Mpl)^2\right].
\end{align}
We comment in passing on another potentially relevant scale for this problem which is the Compton scale of the nucleated brane $r_\mathrm{C} \sim T^{-1/3}$. In the (semi-)classical regime, we expect that the size of the nucleated brane is much larger than its Compton wavelength, i.e. $r_* \gtrsim r_\mathrm{C}$ which gives the same inequality as in eq.~\eqref{eq:exponentialSuppression}. This is the same behaviour as one gets in the original FL bound showing that the conjecture really lies on the boundary of the regime of validity for the nucleation rate in eq.~\eqref{d2rate}. This is of course expected since we get the bound by setting the instanton bounce action $I_\mathrm{B} \sim \mathcal{O}(1)$, which is indeed the boundary of the regime of validity.

\subsection{Scaling arguments for 2-branes in 4 dimensions} 
In the case of the 2-brane, we were able to solve the Euclidean equations of motion explicitly and obtain the nucleation rate. However, for the FL bound in eq.~\eqref{eq:FL2brane}, only the scaling of the exponent was important. This can be obtained without having the explicit solution to the equations of motion. In this section, we review this for the 2-brane and then generalize the argument to arbitrary $p$-branes in arbitrary dimensions $d$. 

As in the previous section, we work in the thin wall limit in which the wall thickness is negligible compared to the bubble size, for which the characteristic length scale is $r_*$. We should look at the scaling behaviour of the action with respect to this parameter. Let us schematically rewrite eq.~\eqref{d2actionEuclidean} in a language more amenable to scaling arguments:
\begin{equation}\label{d2actionforms}
    I_{\mathrm{E}} \sim T\int_{\mathrm{WV}} \star 1+ig\int_{\mathrm{WV}}  A\wedge F^{(\mathrm{B})}+\int_{\mathrm{WV}}  F^{(\mathrm{B})} \wedge \star F^{(\mathrm{B})} + \int_{\mathrm{Bulk}}dA\wedge \star dA,
\end{equation}
where once again we disregard numerical factors for simplicity.

Since we neglect the back-reaction of the bulk electromagnetic field, let us ignore the final term since it will cancel in the evaluation of the bounce action. The equation of motion of the brane gauge field is
\begin{equation}
    d\star F^{(\mathrm{B})} \sim ig\ dA.
\end{equation}
Knowing that $dA\sim E$, i.e. the background electric field, we infer the scaling behavior of the brane field strength $F^{(\mathrm{B})}\sim gE$. Putting this term in eq.~\eqref{d2actionforms}, and integrating over the bubble worldvolume, we find
\begin{equation}\label{bouncesim}
    I_{\mathrm{E}}\sim TR^3+(gE)^2 R^5,
\end{equation}
where the power law behavior for the second term is fixed by dimensional analysis. We once more obtain, in agreement with eq.~\eqref{r_*}, $r_*\sim T^{1/2}/gE$ by minimizing the above action with respect to $R$. This result implies again that the bounce action of eq.~\eqref{bouncesim} scales as $I_{\mathrm{B}}\sim T^{5/2}/(gE)^3$.

We reiterate that such scaling arguments cannot give explicit solutions as in section \ref{sec:2branes}, and they assume such a solution exists. Nevertheless, a major advantage is that one does not have to solve the full equations to obtain the parametric form of the bounce action. In higher dimensions we are not able to solve for the bounce explicitly and we resort to these scaling arguments instead. 

\subsection{Generalizations to $p$-branes in $d$ dimensions}\label{sec:scalingpind}

The methods we have outlined above are can be easily extended to analogous problems in higher dimensions. In all cases, we assume the dimension of the brane can vary, i.e. we allow $p$-branes with $p\leq d-1$ in $d$ dimensions. In all subsequent examples, we continue to assume only a uniform background electric field $E$. Since we will also need to generalize the Wess-Zumino coupling, we will consider the cases of even $p$ and odd $p$ separately. 

\subsubsection{Even $p$-branes}\label{sec:evenpscaling}
For $p$ even, the world volume of the brane is $p+1$ dimensional and requires a $(p+1)$-form for integration in the action. Since we want to consider a coupling of the brane to the bulk 1-form gauge field, we need to make up the form degree by taking the wedge product of $A$ with a $p$-form field (or product of fields). Since $p$ is even, a natural choice\footnote{For example, D-branes of string theory have such couplings in their world volume action as we will discuss further in section \ref{sec:stringtheory}.} is to use branes with a 2-form field strength on their world volume (like $F^{(\mathrm{B})}$ in section \ref{sec:2branes}) and take $p/2$ factors of this field strength. As such, we take the following action which generalizes the coupling in eq.~\eqref{d2actionforms}:
\begin{equation}
\label{eq:evenpAction}
    I_{\mathrm{E}} \sim T_p\int_{\mathrm{WV}} \star 1+ig\int_{\mathrm{WV}}  A\wedge \underbrace{F^{(\mathrm{B})} \wedge \ldots \wedge F^{(\mathrm{B})}}_{\text{$p/2$ factors}} + \int_{\mathrm{WV}}   F^{(\mathrm{B})} \wedge \star F^{(\mathrm{B})} + \int_{\mathrm{Bulk}}dA\wedge \star dA ,
\end{equation}
where the abbreviation $\mathrm{WV}$ again denotes a world-volume integral and we have dropped all $\mathcal{O}(1)$ factors since we are only looking for the scaling behavior of the bounce action. Since dimensional analysis will play an important role in the argument, we quote the dimensions of the fields and couplings in our normalization:
\begin{align}
\label{eq:evenpDims}
[T_p] = p+1 \quad ; \quad [A] = \frac{d-2}{2} \quad ; \quad [F^{(\mathrm{B})}] = \frac{p+1}{2} \quad ; \quad [g] = \frac{4-d}{2} - \frac{p(p-3)}{4}
\end{align}
In what follows, the gauge coupling $g$ will always show up in the combination $gE$ where $E$ is the background electric field. As such, we also quote the dimension of $gE$ and note that it is independent of the spacetime dimension $d$:
\begin{align}
[gE] = 2 - \frac{p(p-3)}{4}.
\end{align}
We are now ready to find the parametric dependence of the bounce action and we start by writing (schematically) the equation of motion of $F^{(\mathrm{B})}$: 
\begin{equation}
	d\star F^{(\mathrm{B})} \sim ig \; dA \wedge F^{(\mathrm{B})} \wedge \ldots \wedge F^{(\mathrm{B})} \sim gE  \underbrace{F^{(\mathrm{B})} \wedge \ldots \wedge F^{(\mathrm{B})}}_{\text{$(p-2)/2$ factors}}.
\end{equation}
Assuming that $F^{(\mathrm{B})}$ is proportional to a power of $gE$, it is easy to use the equation of motion to show that:
\begin{align}
\label{eq:evenpScaling}
F^{(\mathrm{B})} \propto (gE)^{\frac{2}{4-p}}
\end{align}
which already implies that this argument will not work for the case $p=4$. We will comment on this more below but for now let us continue with $p \neq 4$. For these cases, and using the scaling in eq.~\eqref{eq:evenpScaling}, we can evaluate terms in the action of eq.~\eqref{eq:evenpAction}. Since we are ignoring back-reaction of the brane nucleation, we can drop the last term in $I_B$. Moreover, if we solve the $F^{(\mathrm{B})}$ equations of motion, then the second and third terms are parametrically of the same order. We therefore have only two terms given by:
\begin{align}
I_\mathrm{E} \sim T_p R^{p+1} - (gE)^{\frac{4}{4-p}} R^{\frac{4}{4-p} + p+1} 
\end{align}
where $R$ is the characteristic size of the nucleated bubble and the power of $R$ in each term is determined by dimensional analysis. Note that the minus sign is important for the action to have a minimum in $R$. This minimum is at:
\begin{align}
\label{eq:evenpRadius}
R=R_* \sim \frac{T_p^{\frac{4-p}{4}}}{gE}
\end{align}
which gives the bounce action:
\begin{align}
I_\mathrm{B} \sim \frac{T_p^{2 - \frac{p(p-3)}{4}}}{(gE)^{p+1}}.
\end{align}
Note the implication that $F^{\mathrm{(B)}}\sim T_p^{1/2}$ on the instanton solution, as expected from eq.~\eqref{aphisol}, for instance\footnote{The extra factor quadratic in the radius disappears upon raising indices with respect to the induced metric on the brane.} when $p=2$.
As before, we must require that the bounce action is larger than $\mathcal{O}(1)$ or that the nucleated bubble size is too large for our flat space approximation to be valid (i.e. larger than the size of the $S^{d-2}$ in the Nariai solution). Putting these together and using the fact that generic Nariai black holes have an electric field $E \sim M_d^{\frac{d-2}{2}} H$ in $d$ dimensions, we get:
\begin{align}
\label{eq:FLevenp}
T_p^{\frac{1}{p+1} + \frac{4-p}{4}} \gtrsim g M_d^{\frac{d-2}{2}} H \quad \mathrm{or} \quad T_p^{\frac{4-p}{4}} \gtrsim g M_d^{\frac{d-2}{2}} \quad \text{for $p$ even and $p\neq 4$}.
\end{align}
We note that the above bounds are upper bounds on the tensions of branes for $p > 4$ and lower bounds for branes with $p<4$. As such, for higher dimensional branes, the FL bound has a direction similar to the WGC but it is bounding a different coupling. More precisely, the WGC for $p$-branes is a relation between the brane tension and its charge under a $p+1$-form gauge field. Here are discussing a relation between the brane tension and its coupling to a bulk 1-form gauge field (where we make up the form degree using a brane gauge field $F^{(\mathrm{B})}$ as in eq.~\eqref{eq:evenpAction}). Finally, we point out that the Hubble constant $H$ cancels in the second condition. This cancellation arises owing to the simple $E$ dependence in eq.~\eqref{eq:evenpRadius}. It is not clear to us whether this simple dependence is true of higher order corrections.

\subsubsection*{Tension and scaling arguments}
Observe from eq.~\eqref{eq:evenpRadius} that when $p>4$, the radius of the critical bubble \textit{decreases} as the brane tension \textit{increases}. This peculiar pattern is unusual compared to the typical Schwinger pair production of particles (i.e. 0-branes). Indeed, for particle pair production, intuition suggests that as the particle mass is increased at fixed particle charge, a larger bubble is necessary in order for the screening of the electric field to provide the necessary energy for the bubble wall, or the particles; otherwise, there will not be enough of an energy deficit to compensate for the mass of a pair. The subtlety lies in the fact that in the cases that we have, it is not possible to hold the induced electric charge on the brane fixed while increasing the tension. Indeed, since $F^{(\mathrm{B})}\sim T_p^{1/2}$ on the instanton solution, the Wess-Zumino interaction\footnote{It is convenient to interpret the coefficient of the Ramond-Ramond field $A$ in the Wess-Zumino interaction as a current density. Then the current contains $p/2$ factors of $F^\mathrm{(B)}$; see eq.~\eqref{eq:evenpAction}.} indicates that the charge on the brane\footnote{The \textit{total} charge on the brane still vanishes of course, due to the northern and southern hemispheres being oppositely charged.} scales as $T_p^{p/4}$. Therefore, as $T_p$ increases the brane accumulates larger charges on its surface in the presence of a background $E$ field, and hence there is greater electric field screening, whenever $p>0$. As such, a larger bubble is no longer required since the stronger charges on the brane are enough to lead to a significant electric field energy deficit to account for the energy of the brane. It follows that the background electric field can decay through nucleating many small bubbles.\footnote{However, a caveat is that as the back-reaction of the electric and gravitational fields becomes larger, we no longer trust our solutions. Back-reaction effects of extremely compact critical bubbles associated with high tension, should modify these results.} 

To make these arguments more concrete, we proceed with a more detailed explanation. The energy of the nucleated brane is the sum of the energy associated with its tension, as well as the energy of the gauge field living on the brane, both of which are the same order of magnitude owing to the bubble solution extremizing the Euclidean action. On the one hand, the energy of the brane coming from two contributions of similar order is
\begin{align}
    \delta U_{\mathrm{Brane}}\sim \underbrace{R_*^p \times (F^{(\mathrm{B})})^2}_{\substack{\text{brane gauge}\\ \text{field energy}}} \sim  \underbrace{R_*^p \times T_p}_{\substack{\text{brane tension}\\ \text{ energy}}} \sim \frac{T_p^{1+p-\frac{p^2}{4}}}{(gE)^p}.
\end{align}
On the other hand, the \textit{loss} in the electric field energy due to the screening is 
\begin{align}
     |\delta U_{E}|\sim E\times \underbrace{g (F^{\mathrm{(B)}})^{\frac{p}{2}}}_{\substack{\text{change in $E$} \\ \text{in screened volume}}} \times \underbrace{R_*^{p+1}}_{\substack{\text{approximate}\\ \text{screened volume}}}\sim \frac{T_p^{1+p-\frac{p^2}{4}}}{(gE)^p},
\end{align}
which - as expected - matches the energy of the brane, where we once again ignored numerical factors\footnote{For clarity, it is insightful to consider the special case of $p=2$ which we had explicitly solved earlier in section \ref{sec:2branes}. Then $\delta U_{\mathrm{Brane}}\sim T_2^2/(gE)^2$. The volume of the region in which $E$ is screened is that of a sphere $V\sim  R_*^3$ which the brane encloses, so $|\delta U_{E}|\sim gE F^{\mathrm{(B)}} R_*^3\sim T_2^2/(gE)^2$. Moreover, we get back the usual picture of the Schwinger process if we set $p=0$, as expected.}.  We now see why eq.~\eqref{eq:FLevenp} yields upper bounds on the tension when $p>4$: as brane tension increases, the exponent suggests that nucleation rate increases as well. Requiring the nucleation rate to be exponentially slow means that the tension of the brane should not be too large.

\subsubsection{Odd $p$-branes}\label{sec:oddpscaling}
There is a similar story for $p$ odd. In this case, however, it is not possible to get a $(p+1)$-form using one factor\footnote{We're looking for couplings that can charge the $p$-brane electrically under $A$ which is why we take only one such factor.} of $A$ and multiple factors of a 2-form $F^{(\mathrm{B})}$. In order to order to couple the brane to the bulk gauge field $A$, we add an additional axionic degree of freedom $\theta$ on the brane world-volume. This has a one form field strength and allows us to generalize the action of eq.~\eqref{d2actionforms} to the following:
\begin{align}
\label{eq:oddpAction}
I_{\mathrm{E}} &\sim T_p\int_{\mathrm{WV}} \star 1+ig\int_{\mathrm{WV}}  A\wedge d \theta \wedge \underbrace{F^{(\mathrm{B})} \wedge \ldots \wedge F^{(\mathrm{B})}}_{\text{$(p-1)/2$ factors}} + \int_{\mathrm{WV}}   F^{(\mathrm{B})} \wedge \star F^{(\mathrm{B})} \nonumber\\
&+ \int_{\mathrm{WV}} d\theta \wedge \star d \theta + \int_{\mathrm{Bulk}}dA\wedge \star dA 
\end{align}
This axionic degree of freedom can be obtained for example if one starts with a $(p+1)$-brane that wraps a 1-cycle in the internal directions. Then the dimensional reduction of $F^{(\mathrm{B})}$ would contain a zero form gauge field which can play the role of $\theta$ in our case. 

As in the previous section, we start by quoting the mass dimension of the couplings and fields as well as the combination $gE$:
\begin{align}
[d\theta ] = \frac{p+1}{2} \quad ; \quad [g] = \frac{7-2d}{4} - \frac{p(p-2)}{4} \quad ; \quad [gE] = \frac{7}{4} - \frac{p(p-2)}{4}.
\end{align}
The bulk gauge field $A$ and brane field strength $F^{(\mathrm{B})}$ have the same dimensions as in eq.~\eqref{eq:evenpDims}. Next, we turn to the equations of motion of the brane degrees of freedom, of which we now have two. We write these equations in the presence of a background electric field $E$:
\begin{align}
d\star F^{(\mathrm{B})}  &\sim gE \;  d\theta \wedge \underbrace{F^{(\mathrm{B})} \wedge \ldots \wedge F^{(\mathrm{B})}}_{\text{$(p-3)/2$ factors}}\\
d\star d\theta &\sim gE \underbrace{F^{(\mathrm{B})} \wedge \ldots \wedge F^{(\mathrm{B})}}_{\text{$(p-1)/2$ factors}}.
\end{align}
We can again for the scaling of $F^{(\mathrm{B})}$ and $d\theta$ with $gE$ and we find from the equations of motion:
\begin{align}
F^{(\mathrm{B})} \sim d\theta \sim (gE)^{\frac{2}{3-p}}.
\end{align}
Similar to the case of even $p$, we see that we cannot determine the scaling of the field strengths for $p = 3$ but we again postpone this discussion until section \ref{sec:3and4branes}. In this section, we will consider the cases where $p\neq 3$ for which we can evaluate the Euclidean action and minimise it to get 
\begin{align}
\label{eq:oddpRadius}
R_* \sim \frac{T_p^{\frac{3-p}{4}}}{gE}
\end{align}
and a bounce action:
\begin{align}
I_\mathrm{B} \sim \frac{T_p^{\frac{7}{4} - \frac{p(p-2)}{4}}}{(gE)^{p+1}}.
\end{align}
Given these values, the FL bound becomes:
\begin{align}
\label{eq:FLoddp}
T_p^{\frac{1}{p+1} + \frac{3-p}{4}} \gtrsim g M_d^{\frac{d-2}{2}} H \quad \mathrm{or} \quad T_p^{\frac{3-p}{4}} \gtrsim g M_d^{\frac{d-2}{2}} \quad \text{for $p$ odd and $p\neq 3$}.
\end{align}
Once more, these inequalities are upper bounds on the tensions of branes for $p > 3$ and lower bounds on the tensions of branes for $p < 3$. On a passing note, we mention that this behavior of tension is akin to the case of even $p$ branes, which we have already explained. Indeed, since $d\theta\sim T_p^{1/2}$ as well - in addition to $F^{\mathrm{(B)}}$ - on the Euclidean solution, for larger $p$, the Wess-Zumino coupling makes the induced charge on the brane increase with increasing tension. 

\subsection{The 3- and 4-brane exceptions}
\label{sec:3and4branes}

We have so far not discussed D3 and D4 branes. These are somewhat more subtle. Take for instance the action for a D4 brane:
\begin{align}\label{d4action}
     I_{\mathrm{E}} \sim & T\int_{\mathrm{WV}} \star 1+ig\int_{\mathrm{WV}}  A \wedge F^{(\mathrm{B})} \wedge F^{(\mathrm{B})}+\int_{\mathrm{WV}}  F^{(\mathrm{B})} \wedge \star F^{(\mathrm{B})} \\ &  + \int_{\mathrm{Bulk}}dA\wedge \star dA.
\end{align}
Varying with respect to $A^{(\mathrm{B})}$ shows that 
\begin{equation}
    d\star F^{(\mathrm{B})} \sim ig\ dA\wedge F^{(\mathrm{B})}.
\end{equation}
The complication lies in the fact that the equation of motion possesses an accidental scaling symmetry, in the sense that whatever solution we may obtain for $A^{(\mathrm{B})}$, linearity guarantees that $\lambda A^{(\mathrm{B})}$ is also a solution, for some number $\lambda$. The action does not have this symmetry. This feature is reminiscent of the small amplitude simple pendulum system, whose action is
\begin{equation}
    I=\int dt \Big(\frac{1}{2}\dot\theta^2-\frac{1}{2} \omega^2 \theta^2 \Big).
\end{equation}
The equation of motion is that of a simple harmonic oscillator:
\begin{equation}
\ddot{\theta}+\omega^2 \theta=0.
\end{equation}
The implication of this dilatation symmetry ($\theta\rightarrow \lambda \theta$) at the level of the equation of motion, albeit not the action, is that the frequency of oscillation $\omega$ does not fix the amplitude, which remains an independent parameter. In the case of D4 branes, the accidental symmetry implies that the critical size of the bubble is not fixed by the terms in the action so far. We must include an extra piece in the action that breaks the scaling symmetry at the level of the equation of motion. The same issue occurs with D3 branes: there is a simultaneous scaling of the brane gauge field as well as the brane scalar field shows an accidental symmetry at the level of the equation of motion.

\subsection{Other brane charges}
\label{sec:OtherBraneCharges}

Throughout the previous discussion, we considered branes whose world volume action contains terms of the form shown in eqs.~\eqref{eq:evenpAction} and~\eqref{eq:oddpAction}. However, it is natural for a $p$-brane to be charged under a $(p+1)$-form gauge field and this should also be included in the WV action. In fact, all D-branes of string theory are charged in this way and understanding this coupling will be important when we try to verify our FL bound using the tensions and couplings of D-branes. As such we add to the action terms of the form:
\begin{align}
\label{eq:braneChargeAction}
S \supset \gamma\int_{\rm WV} C_{p+1} + \int_{\rm Bulk} dC_{p+1} \wedge \star dC_{p+1}.
\end{align}
In the presence of this coupling, a nucleated $p$-brane will back-react on the field strength $dC_{p+1}$ and this will also contribute to our estimate of the bounce action\footnote{Note that higher form gauge fields can be used to dynamically reduce other quantitites such as the cosmological constant, see for example \cite{Brown:1988kg, Feng:2000if}.}. In fact, if the coefficient $\gamma$ is too large, it may overwhelm the contribution of the coupling to the electromagnetic gauge field $A$. In this section, we will derive the condition under which this contribution is negligible so that one may trust the FL bounds of eq.~\eqref{eq:FL2brane}, eq.~\eqref{eq:FLevenp}, and eq.~\eqref{eq:FLoddp}. 

As before we will use scaling arguments and dimensional analysis to estimate the contribution of eq.~\eqref{eq:braneChargeAction} to the bounce action. As such, we begin by quoting the mass dimension of the constant $\gamma$:
\begin{align}
[\gamma] = p + 2 - \frac{d}{2}.
\end{align}
Next, we inspect the equation of motion of $C_{p+1}$:
\begin{align}
d\star dC_{p+1} \sim \gamma \delta({\rm WV})
\end{align}
where $\delta_{\rm WV}$ is a delta-function $(d-p-1)$-form that picks out the brane world-volume. It is clear from this equation that the solution will have $C_{p+1} \propto \gamma$. We may then use dimensional analysis to estimate the contribution of eq.~\eqref{eq:braneChargeAction}:
\begin{align}
\Delta S \sim \gamma^2 R_*^{2p + 4 - d}.
\end{align}
We must require this contribution to the action to be much less than that of the other terms, all of which are of the order of $T_p R_*^{p+1}$. We therefore get, after using the expressions for $R_*$ from eqs.~\eqref{eq:evenpRadius} and~\eqref{eq:oddpRadius}:
\begin{align}
\label{eq:evenbraneChargeCondition}
\gamma^2 &\lesssim \frac{T_p^{1 + \frac{1}{4}(4-p)(d-p-3)}}{(gE)^{d-p-3}} \quad ; \quad \text{for $p$ even and $p\neq 4$}\\
\label{eq:oddbraneChargeCondition}
\gamma^2 &\lesssim \frac{T_p^{1 + \frac{1}{4}(3-p)(d-p-3)}}{(gE)^{d-p-3}} \quad ; \quad \text{for $p$ odd and $p\neq 3$}
\end{align}

\subsection{Higher form bulk fields and back-reaction}\label{sec:back-reaction}
To conclude this section, let us mention that till now we have been considering a one-form gauge field on the brane in our bubble configurations. While this choice may give self-consistent solutions in their own right, we would also like to check that our solutions remain valid even if gauge fields in the bulk, of higher form degree, exist and couple to the brane. This investigation will be particularly useful when we compare our analysis to branes in string theory, where such higher form fields are ubiquitous. In the case of $2$-branes, this complication does not matter as the brane dimension is too low to allow it to couple to higher form bulk fields through a Wess-Zumino interaction. But now that we have also explored higher $p$-branes, let us verify that solutions remain self-consistent even in the presence of these higher form bulk fields that couple to the brane. In other words, we show that these higher form fields remain small, if initially small, and do not affect our solution. We emphasize that we will focus on the limit of \textit{weak coupling}, which is to say that $g\rightarrow 0$, $E \rightarrow \infty$, while holding $gE$ fixed. We will not discuss any gravitational back-reaction.

\subsubsection*{Even $p$-branes}

Starting with a concrete example, consider first the special case of adding a bulk $3$- form gauge field in the action with Wess-Zumino type coupling. Take the action:
\begin{align}\label{d4action}
     I_{\mathrm{E}} \sim & T_p\int_{\mathrm{WV}} \star 1+ig_{(1)}\int_{\mathrm{WV}}  A_{(1)} \wedge \underbrace{F^{(\mathrm{B})} \wedge \ldots \wedge F^{(\mathrm{B})}}_{\text{$p/2$ factors}}+ ig_{(3)}\int_{\mathrm{WV}}  A_{(3)} \wedge \underbrace{F^{(\mathrm{B})} \wedge \ldots \wedge F^{(\mathrm{B})}}_{\text{$(p-2)/2$ factors}}  \\ &   +\int_{\mathrm{WV}}  F^{(\mathrm{B})} \wedge \star F^{(\mathrm{B})}  + \int_{\mathrm{Bulk}}dA_{(1)}\wedge \star dA_{(1)}+\int_{\mathrm{Bulk}}dA_{(3)}\wedge \star dA_{(3)}.\nonumber
\end{align}
Here, we have indexed both the fields and corresponding Wess-Zumino couplings for clarity. Varying the action gives the following equations of motion:
\begin{align}\label{formeq1}
    d\star F^{\mathrm{(B)}}\sim g_{(1)} dA_{(1)}\wedge \underbrace{F^{(\mathrm{B})} \wedge \ldots \wedge F^{(\mathrm{B})}}_{\text{$(p-2)/2$ factors}}+g_{(3)} dA_{(3)}\wedge \underbrace{F^{(\mathrm{B})} \wedge \ldots \wedge F^{(\mathrm{B})}}_{\text{$(p-4)/2$ factors}},
\end{align}

\begin{align}\label{formeq2}
    d\star dA_{(1)}\sim g_{(1)} \underbrace{F^{(\mathrm{B})} \wedge \ldots \wedge F^{(\mathrm{B})}}_{\text{$p/2$ factors}}\delta(\mathrm{WV}),
\end{align}

\begin{align}\label{formeq3}
    d\star dA_{(3)} \sim g_{(3)} \underbrace{F^{(\mathrm{B})} \wedge \ldots \wedge F^{(\mathrm{B})}}_{\text{$(p-2)/2$ factors}}\delta(\mathrm{WV}).
\end{align}
Consider first a vanishing $dA_{(3)}$ background (the $dA_{(1)}\sim E$ background is still non-vanishing of course). Then it follows from scaling arguments that $F^{\mathrm{(B)}}\sim (g_{(1)}E)^{\frac{2}{4-p}}$. For this result to remain self-consistent, we must check that the back-reaction of $dA_{(3)}$ is not too large. Indeed, we see that $\delta dA_{(3)}\sim g_{(3)} (g_{(1)}E)^{\frac{p-2}{4-p}}$, which remains small when the other couplings are also weak, ie. in the limit that $g_{(3)}\rightarrow 0$. Our equations would \textit{not} have been consistent if we had found, for instance, that as $g_{(1)}, g_{(3)}\rightarrow 0$, the change in $dA_{(3)}$ became large or even singular (again, $g_{(1)}E$ is held fixed in the weak coupling regime). 

One can repeat this argument with higher form bulk fields, yielding similar results. Explicitly, if we had considered a bulk $k$-form field, with $k$ odd, then we would append to the action the following:
\begin{equation}
    g_{(k)} \int_{\mathrm{WV}}  A_{(k)} \wedge \underbrace{F^{(\mathrm{B})} \wedge \ldots \wedge F^{(\mathrm{B})}}_{\text{$(p-k+1)/2$ factors}}+\int_{\mathrm{Bulk}}dA_{(k)}\wedge \star dA_{(k)}.
\end{equation}
Arguing as before, the back-reaction on the $dA_{(k)}$ field (which was initially vanishing) is $\delta dA_{(k)}\sim g_{(k)}(g_{(1)}E)^{\frac{p-k+1}{4-p}}$, which - for fixed $g_{(1)}E$ - becomes vanishingly small as $g_{(k)}\rightarrow 0$. It is also clear that if one were to include multiple higher form bulk gauge fields, the same conclusion would follow: eq.~\eqref{formeq1} would be incremented by additional expected terms, analogs of eq.~\eqref{formeq3} would emerge, and the preceding argument would carry through as expected.

\subsubsection*{Odd $p$-branes}
A similar result follows in the case of odd $p$-branes. Indeed, say 
\begin{align}
I_{\mathrm{E}} &\sim T_p\int_{\mathrm{WV}} \star 1+ig_{(1)}\int_{\mathrm{WV}}  A_{(1)}\wedge d\theta \wedge \underbrace{F^{(\mathrm{B})} \wedge \ldots \wedge F^{(\mathrm{B})}}_{\text{$(p-1)/2$ factors}} + \int_{\mathrm{WV}}   F^{(\mathrm{B})} \wedge \star F^{(\mathrm{B})} \\
&+ \int_{\mathrm{WV}} d\theta \wedge \star d \theta + \int_{\mathrm{Bulk}}dA_{(1)}\wedge \star dA_{(1)} \nonumber \\&+ig_{(k)}\int_{\mathrm{WV}}A_{(k)}\wedge d\theta \wedge \underbrace{F^{(\mathrm{B})} \wedge \ldots \wedge F^{(\mathrm{B})}}_{\text{$(p-k)/2$ factors}}+\int_{\mathrm{Bulk}}dA_{(k)}\wedge \star dA_{(k)},\nonumber
\end{align}
where the last line modifies eq.~\eqref{eq:oddpAction} with the inclusion of an extra (odd) $k$-form bulk field $A_{(k)}$; once more, we index the coupling parameters and bulk gauge fields for clarity. Repeating the same arguments above, as well as section \ref{sec:oddpscaling}, we see that the back-reaction of the $dA_{(k)}$ field goes as $\delta  dA_{(k)}\sim g_{(k)}(g_{(1)}E)^{\frac{4+(p-k)(3-p)}{2(3-p)}}$. This expression too becomes small as $g_{(k)}\rightarrow 0$, with $g_{(1)}E$ held fixed, ensuring the back-reaction of the $dA_{(k)}$ field remains small in the weak coupling limit. In short, our solutions in section \ref{sec:scalingpind} continue to remain valid at weak coupling.

\section{Branes without world-volume gauge fields}
\label{sec:FLforBranesnoWVGF}

The branes studied in the previous section were inspired by the D-branes present in string theory. These objects feature world-volume degrees of freedom originating from open strings ending on these branes. These degrees of freedom include world-volume gauge fields that played an important role in our analysis. In particular, the world-volume gauge fields acquired a non-trivial classical profile in the solutions that we studied and this allowed the brane to couple to the bulk gauge fields via the Wess-Zumino terms in eq.~\eqref{eq:evenpAction} and eq.~\eqref{eq:oddpAction}. Due to this coupling and the non-trivial gauge field profile, the branes obtain an electric charge and are able to screen the bulk electric fields. More generally, we may imagine extended sources without such world-volume gauge fields. These could represent solitonic objects in quantum field theory or unknown objects yet to be discovered in string theory. An axion domain wall provides an example of these objects (and we will say more about them below) since it couples to bulk 1-form gauge fields without having world-volume-localised gauge fields. In this section, we will study branes of this type which are inspired by axion domain walls. 

The aim is to derive FL bounds on the properties of 2-dimensional branes without world-volume gauge fields. The general setup we study is the same as the one we considered in section~\ref{sec:FLforBranes}. Namely, we will consider electrodynamics coupled to gravity in de Sitter space, eq.~\eqref{eq:EinsteinMaxwell}. In addition, we will assume the theory also contains 2+1 dimensional branes with worldvolume action given by:
\begin{align}
	\label{eq:braneWV}
	I_{\rm 2-brane} = \int_{\mathrm{WV}} \Bigg( T_2 *1 - \frac{g^2}{4\pi} A \wedge F \Bigg),
\end{align}
where $T_2$ is the tension of the 2-brane, $g$ is gauge coupling of the theory of electrodynamics. As mentioned above, this coupling to bulk gauge fields resembles the axion domain wall coupling. The axion domain wall is the special case where the tension $T_2$ is also determined by the axion mass and decay constant and is given by $T_2\sim mf^2$. Nonetheless, the bound we derive in this section will not apply to axion domain walls as we explain below. Branes with similar couplings may also be constructed within string theory in the context of the quantum Hall effect~\cite{Bergman:2004mv}.

In previous sections, we studied electrically charged Nariai solutions of the theory in Eq.~\eqref{eq:EinsteinMaxwell}. This theory also has dyonic Nariai black hole solutions~\cite{Lee:1991jw,Balakin:2017nbg} which will be the focus of this section. Analogous to their electrically charged counterparts, these Nariai solutions are $dS_2 \times S^2$ spacetimes with constant parallel electric and magnetic fields stretching between the black hole and de Sitter horizons. More details on these solutions are presented in appendices~\ref{sec:appendixDyonicBHs} and~\ref{sec:appendixDyonicwAxions}. In a previous paper~\cite{Hassan:2024nbl}, we showed that such a (flat space) field configuration is unstable to nucleating axion domain walls. More generally, the same process allows for the nucleation of any 2-brane with the action given in eq.~\eqref{eq:braneWV}. This nucleation process can screen the electric and magnetic fields and is also expected to happen in curved spacetimes such as the dyonic Nariai spacetime. By the same argument as the original FL conjecture, we will use the flat space result to estimate the decay rate since the screening happens locally and thus essentially in flat space. We will describe the nucleation of the 2-brane in section~\ref{sec:domainWallNucleation}. The most important aspect of this process is the decay rate which is given by:
\begin{align}
\label{eq:decayRate}
\Gamma/V \sim \exp \left[-c \frac{T_2^4}{(g^2 E B)^3} \right].
\end{align}
where $E$ and $B$ are the magnitudes of the parallel background electric and magnetic fields and $c$ is an order one constant that is known in examples~\cite{Hassan:2024nbl}. This rate is exponentially suppressed for $E$ and $B$ fields with magnitude much smaller than the tension of 2-brane $T_2$. The bound we derive follows from the observation that Nariai black holes have $E \sim B \sim \Mpl H$. As such, this screening process can be fast for $T_2$ sufficiently small. Demanding that this process is exponentially suppressed gives the constraint:
\begin{align}
\label{eq:braneFL}
T_2 \gtrsim (g \Mpl H)^{3/2}
\end{align}
which, modulo a loophole we outline below, gives a bound on the wall tension. 

We expect the bound in eq.~\eqref{eq:braneFL} to apply whenever the requisite dyonic Nariai black holes exist. In particular, these black holes have the large electric and magnetic fields $E \sim B \sim \Mpl H$. We can thus evade this bound by ensuring that all classically stable Nariai black holes do not have electromagnetic fields comparable to $\Mpl H$. This is easy to do by including a light axion in the theory which can shift to classically screen the electric field. This is the loophole we discuss in section~\ref{sec:loopholes}, specifically the discussion around eq.~\eqref{WittenChange}. In this case, one cannot apply the FL argument and the constraint eq.~\eqref{eq:braneFL} does not hold\footnote{We are deeply grateful to Prateek Agrawal for pointing this out in a conversation which he may no longer remember.}. In particular, this is the reason that our constraint does not hold for axion domain walls, as we will describe in more detail in section~\ref{sec:loopholes}.

\subsection{Brane nucleation}
\label{sec:domainWallNucleation}
In this section, we will briefly describe the nucleation of 2-branes with action eq.~\eqref{eq:braneWV} from parallel electric and magnetic fields. As mentioned previously, for our purposes we will consider the nucleation of branes in flat space and ignore gravitational effects. 

We will consider a setup where we have parallel electric and magnetic fields in flat space and work in the no back-reaction limit. In this limit, we take the coupling $g \rightarrow 0$ and make the electromagnetic fields large so as to keep the combination $g^2 F \wedge F$ fixed. This is a convenient limit because the nucleation of a brane does not alter the surrounding electromagnetic fields. These back-reaction effects are interesting and have been studied in~\cite{Hassan:2024nbl} for axion domain walls but are unimportant for our discussion here.

The process we are interested in involves nucleating a brane from the vacuum that expands and screens the electromagnetic fields due to the $A\wedge F$ coupling. In order to describe the nucleation process, and in particular the nucleation rate, we proceed along the lines of \cite{brown2018schwinger} and work with the worldvolume action in eq.~\eqref{eq:braneWV}. As usual, the prescription \cite{Coleman:1977py, Coleman:1977th, Coleman:1985rnk, Callan:1977pt} is to first Wick rotate the action to Euclidean signature and extremize this action subject to suitable boundary conditions. The nucleation rate is then:
\begin{align}
\Gamma/V \sim \exp{(-I_{\mathrm{B}})}
\end{align}
where $I_{\mathrm{B}}$ is the Euclidean bounce action. We will assume that the brane is nucleated in an initially spherical configuration (i.e. preserving the full $O(4)$ symmetry of Euclidean space) which is appropriate in the no back-reaction limit.

We begin by quoting the action of eq.~\eqref{eq:braneWV} written in Euclidean signature:
\begin{align}
	\label{eq:EuclideanWV}
	I_{\rm E} = \int_{\mathrm{WV}}\Bigg( T_2 *1 - i \frac{g^2}{4\pi} A \wedge F \Bigg).
\end{align}
We orient the background electric and magnetic fields along the $+z$-axis. As mentioned above, we will assume that the brane is nucleated in a spherical configuration. In this case, the action above depends only on the sphere radius which is determined by demanding that the action is extremized. The two terms in the action can be easily evaluated assuming the spherical ansatz. The first term is simply proportional to the brane volume (i.e. the volume of an $S^3$ with radius $r$). The second term can be calculated using Stoke's theorem to convert the integral over the $S^3$ surface to an integral over the interior of brane (i.e. the ball $B^4$ such that $\partial B^4 = S^3$). This gives:
\begin{align}
I_{\rm E} \sim T_2 r^3 - g^2 E B  r^4
\end{align}
where we have neglected $\mathcal{O}(1)$ factors. This is similar to our approach in Sec.~\ref{sec:FLforBranes} and is justified since we will in any case not be able to fix various $\mathcal{O}(1)$ factors below. This action is extremized for:
\begin{align}
r_* \sim \frac{T_2}{g^2 E B}
\end{align}
and gives a bounce action:
\begin{align}
I_{\mathrm{B}} \sim \frac{T_2^4}{(g^2 E B)^3}.
\end{align}

We can also describe this tunnelling process where back-reaction is not negligible. In this case, the brane has effective charge and current densities that (anti-)screen the electromagnetic fields in a complicated way~\cite{Hassan:2024nbl}.

\subsection{Prerequisites and loopholes}
\label{sec:loopholes}

If we use the FL logic to demand that the process derived in the previous section is slow, we get the bound:
\begin{align}
\label{eq:braneFL}
T_2 \gtrsim (g\Mpl H)^{3/2}
\end{align}
on the tension of the brane in terms of cosmological parameters. In this subsection, we discuss two conditions that have to be satisfied for the above bound to apply. We will find that the first condition is met automatically while the second provides a loophole that involves other particles (axions) in the theory.  

Let us start with the first condition. As mentioned previously, we will rely on the flat space nucleation rate eq.~\eqref{eq:decayRate} and as such we have to ensure that the flat space approximation is sensible. This is the case if bubble nucleation happens on scales shorter than the typical curvature scales of the spacetime we are considering. A good proxy of this curvature scale is the sphere radius given in Eq.~\eqref{eq:S2radius} (or~\eqref{eq:S2radiuswithAxion} in the presence of an axion as we discuss below) in units of $\sqrt{\Lambda/3}$. It is easy to see that for typical $\mathcal{O}(1)$ charges, the sphere radius is 
\begin{align}
\tilde{r}_{\rm Nariai} \sim \sqrt{\Lambda} \sim H
\end{align}
where we are using $H$ for the Hubble rate of the 4D $dS$ spacetime without the black hole. As such, the flat space approximation holds when the nucleation radius satisfies:
\begin{align}
r_* \lesssim H^{-1}.
\end{align}
This condition parallels the one we found in sections~\ref{sec:evenpscaling} and~\ref{sec:oddpscaling} when demanding that the bubble fits in the Nariai spacetime and indeed it takes exactly the same form. Substituting the expression for $r_*$, we find:
\begin{align}
    T_2 \lesssim g^2 \Mpl^2 H.
\end{align}
There are now two cases to consider. In the first case, the inequality above is satisfied and we can apply the FL argument and get the bound eq.~\eqref{eq:braneFL}. In the second case, this inequality is violated which means:
\begin{align}
T_2 &\gtrsim g^2 \Mpl^2 H \sim (g \Mpl H)^{3/2} \left[\frac{(g \Mpl H)^{1/2}}{H}\right].
\end{align}
This is in fact stronger than the inequality eq.~\eqref{eq:braneFL} we are trying to derive if the ratio in square brackets is larger than unity. This is indeed the case as we now argue. First, since this theory contains a $U(1)$ gauge group, the WGC demands the existence of a charged particle with mass $m_{\rm WGC} \lesssim g \Mpl$. On the other hand, the original FL bound in eq.~\eqref{eq:originalFL} demands that $m_{\rm WGC}^2 \gtrsim g \Mpl H$. Eliminating $m_{\rm WGC}$ between these two inequalities gives $g \Mpl \gtrsim H$ which directly implies that the ratio in square brackets is bigger than unity. As such, the bound eq.~\eqref{eq:braneFL} is satisfied in either case.

We now turn to a discussion of the loophole we mentioned earlier. In order to derive the bound in eq.~\eqref{eq:braneFL}, we started with a classical Nariai solution with large electric and magnetic fields $E \sim B \sim \Mpl H$. These classical solutions are described in Appendix~\ref{sec:appendixDyonicBHs}. In the presence of an axion that couples to the bulk $F \wedge F$, the electric field of these dyonic Nariai solutions is given by
\begin{align}\label{WittenChange}
    E = \frac{1}{\tilde{r}^2}\left(\tilde{Q}_E - K \frac{\alpha \theta_*}{\pi} \tilde{Q}_M\right).
\end{align}
where the term proportional to the magnetic charge is due to the Witten effect~\cite{Witten:1979ey}. By adjusting the axion vev in the second term, one can (partially) cancel the electric field of these Nariai black holes. If the axion is sufficiently massive, this adjustment is not energetically favorable and there are classical solutions with large electric fields, i.e. where the cancellation is irrelevant. However, for very light axions, the axion field value can adjust to affect a large cancellation in the electric field. Nariai black hole solutions in the presence of an axion field are described in Appendix~\ref{sec:appendixDyonicwAxions}. Taking this effect into account we find that the electric field, in the presence of a light axion, is reduced to:
\begin{align}
\label{eq:dyonicElectricFieldWithAxion}
    E \sim \Mpl H \times \min\left\{1, \left(\frac{m f}{g \Mpl H}\right)^2 \right\}
\end{align}
and vanishes in the limit of an exactly massless axion. Therefore, the presence of a light axion with a coupling to the bulk gauge field in question means that classical dyonic Nariai solutions have a small electric field and cannot be used for the FL argument. We emphasize that this only applies to dyonic black holes because the contribution that cancels the large electric field is proportional to the magnetic charge of the black hole (as is usual in the Witten effect). Therefore, the bound eq.~\eqref{eq:braneFL} on the tension of branes like those in eq.~\eqref{eq:braneWV} only applies if no such light axion exists in the theory.

Finally, we turn to a discussion of eq.~\eqref{eq:braneFL} and why it does not apply to axion domain walls. These domain walls with a coupling like in eq.~\eqref{eq:braneWV}, automatically imply the presence of an axionic degree of freedom in the theory which couples to $F \wedge F$. As we showed above (and in appendix~\ref{sec:appendixDyonicwAxions}), this axion affects the electric field of a dyonic Nariai black hole and can classically screen this field so that it is proportional to the axion mass (see eq.~\eqref{eq:dyonicElectricFieldWithAxion}). In order to derive a bound on the tension of axion domain walls, we need to ensure that the axion is massive enough which means:
\begin{align}
    m \gtrsim \frac{g \Mpl H}{f}.
\end{align}
This is a prerequisite for the existence of dyonic Nariai black holes with large electric and magnetic fields. It is easy to see that this condition (along with $f > m$) already implies that $mf^2 \gtrsim (g \Mpl H)^{3/2}$ which is the FL bound we would get for axion domain walls. As such, we do not learn anything new from the FL argument as the bound is automatically implied by the assumption that dyonic Nariai black holes exist with large electromagnetic fields. 

\section{Comparison with string theory}\label{sec:stringtheory}

The bounds we found in section~\ref{sec:FLforBranes} apply to any branes in dS space that have a world-volume action of the form we assumed. This form of the action is similar (and in fact inspired by) the action for D-branes in Type II strings. The question we want to address in this section is whether the known tensions and couplings of these branes satisfy the inequalities we found in section~\ref{sec:FLforBranes}. More concretely, we imagine a compactification of string theory that gives a $d$-dimensional dS space, along the lines of KKLT or LVS for $d=4$ for example \cite{Kachru:2003aw,Cicoli:2008va,Crino:2020qwk}. At this point, we will not have in mind any such explicit construction but it would be interesting to repeat these checks within the context of an explicit construction. That said, and even without an explicit construction, we will always assume that we are working with large internal volumes and weak\footnote{In fact, as we will see from sections~\ref{sec:stringbackreaction} and~\ref{sec:rrcharges}, we will need to go to a parametrically weak coupling region. It is possible that no dS vacua exist~\cite{Dine:1985he,Grimm:2019ixq,Rudelius:2022gbz} in this region, nonetheless this is the limit in which we can trust our calculations so we will use it for checking the FL inequalities.} couplings so that tree level supergravity remains a valid approximation. In this limit, the D-branes of Type II will still be present in the lower dimensional dS theory\footnote{Their properties may be modified by small effects of $\mathcal{O}(H)$ but we will ignore these modifications for the arguments in this section.}. Since our brane FL derivation is general we should expect that these D-branes satisfy the bounds we found earlier. This is a non-trivial check since the bounds were derived \textit{without} assuming string theory and moreover, there are various inequalities that have to be satisfied across multiple dimensions and for any dimension brane. Pleasingly, we will see that string theory satisfies all these conditions which gives further evidence for the FL argument. 

In order to check this agreement between the conditions we have found and the properties of string theory branes, we need to express the quantities that appear in the FL bounds in terms of string theory variables. To that end, we will examine the Type II supergravity and brane DBI and Chern-Simons (CS) actions and read off the relevant quantities. That said, since we are not really constructing a lower dimensional dS space, we will not be able to express the Hubble rate $H$ in terms of string theory variables. Instead we will simply assume that this takes any value subject to the condition $H < M_s$ where $M_s$ is the string scale. This condition ensures that we have a large classical and non-stringy dS space as was assumed in section~\ref{sec:FLforBranes} for our derivation of the brane FL bounds. The assumption that $H$ can scan a wide range of values is consistent with the picture one has in mind when discussing a landscape of dS vacua~\cite{Susskind:2003kw,Kachru:2003aw,Douglas:2003um,Douglas:2006es,Denef:2007pq,Linde:2015edk} for instance.

\subsection{Preliminaries}\label{stringpreliminaries}
We start by recalling the DBI and CS actions for a D$p$-brane (see for example~\cite{Polchinski:1998rq,Polchinski:1998rr}) which are typically written with the following normalizations:\footnote{In this section, as in the previous ones, we will ignore numerical $\mathcal{O}(1)$ factors. It is not necessary to keep these factors since they will not play a role in any of our checks given that we are unable to fix the $\mathcal{O}(1)$ numbers in our FL inequalities (see section \ref{sec:FLforBranes}).}
\begin{align}
    \label{eq:DBIaction}
    I^p_{\rm DBI} = \frac{M_s^{p+1}}{g_s}\int_{\rm WV} \sqrt{-g_{\rm ind.}} + \frac{M_s^{p-3}}{g_s}\int_{\rm WV} \sqrt{-g_{\rm ind.}}|F^{(\rm B)}|^2
\end{align}
where $g_s$ is the 10-dimensional string coupling (equal to $e^\phi$ where $\phi$ is the dilaton). In writing this action we assumed a vanishing $B$-field and used $F^{(\rm B)}$ for the world-volume field strength tensor to mimic our notation in section \ref{sec:FLforBranes}. First, we can directly read off the brane tension:
\begin{align}
\label{eq:pbraneTension}
    T_p = \frac{M_s^{p+1}}{g_s},
\end{align}
which is simply the coefficient of the first term. We mention that wrapping a $p$-brane on an internal $n$-cycle of volume $V_n$ gives a lower dimensional $(p-n)$-brane whose tension is simply $T_p V_n$. This is larger than $T_{p-n}$ given our assumption that all internal volumes are larger than the string scale. Second, we see that in this action the kinetic term for $F^{(\rm B)}$ is not canonically normalized. To compare to eq.~\eqref{eq:evenpAction} and eq.~\eqref{eq:oddpAction} we have to canonically normalize it by rescaling:
\begin{align}
    \label{eq:FBscaling}
    F^{(\rm B)} \rightarrow \left(\frac{g_s}{M_s^{(p-3)} V_n}\right)^{1/2} F^{(\rm B)},
\end{align}
where again we assumed that the $p$-brane may wrap an internal $n$-cycle to give a $(p-n)$-brane. In writing the rescaling in this way, it is understood that $V_0 = 1$ is the case where the brane does not wrap any internal directions. 

Next, we move on to examining the CS action:
\begin{align}
    \label{eq:CSaction}
    I^p_{\rm CS} = \sum_{0\leq q \leq \frac{p+1}{2}} M_s^{p+1-2q} \int_{\rm WV} C_{p+1-2q} \wedge \left( F^{(\rm B)}\right)^q,
\end{align}
where $C_p$ are the RR $p$-form gauge fields. These couplings are similar to eqs.~\eqref{eq:evenpAction} and~\eqref{eq:oddpAction} and for us the RR forms (suitably reduced) will play the role of the 1-form bulk gauge field $A$. The coefficient of this action will determine the coupling $g$ in the FL bounds after we rescale the RR and $F^{(\rm B)}$ to have a canonical kinetic term. We already saw how to do this for $F^{(\rm B)}$ in eq.~\eqref{eq:FBscaling}. For the RR gauge fields, we need to inspect the Type II action. We will not write this explicitly but simply state that the kinetic terms of these fields can be made canonical by rescaling:
\begin{align}
    \label{eq:RRfieldscaling1}
    C_p \rightarrow \frac{1}{M_s^{4}\sqrt{W_{10-d}}} C_p,
\end{align}
where $W_{10-d}$ is the total volume of the internal manifold. Note that the case $d=10$ is not relevant for us since the vacuum of Type II is not a dS space and we restrict\footnote{Although we do not currently have proposals for constructing dS space with $d > 4$, we will not impose this condition here.} to $d \leq 9$. The volume $W_{10-d}$ is different from $V_n$ which we used earlier to refer to the volume of cycles wrapped by branes. In some of our discussions below, we may wish to reduce the form degree of these RR gauge fields when the brane wraps an internal $n$-cycle. This means that in the lower dimensional theory, we can get a $C_{p-m}$ form, with $m\leq n$, coupled to the unwrapped directions of the brane by performing the integral in~\eqref{eq:CSaction} over $m$ internal directions. In this case, we need to make sure that the kinetic term of this $C_{p-m}$ form is canonically normalized and this involves an additional factor of the interval volume\footnote{This scaling for canonical normalization assumes a geometry that is factorizable. In more complicated manifolds with warping, we may get a different scaling (compare for example the manifolds studied in~\cite{Conlon:2006tq,Svrcek:2006yi}). We will comment more on this scaling in the conclusion (sec.~\ref{sec:conclusion}).}:
\begin{align}
    \label{eq:RRfieldscaling2}
    C_{p-m} \rightarrow \frac{V_m}{M_s^{4}\sqrt{W_{10-d}}} C_{p-m}.
\end{align}
If the $n$-cycle is isotropic then we may write $V_n = L^n$ and $V_m = L^m$ for some length scale $L$ (and as usual up to $\mathcal{O}(1)$ factors). Although we will use this scaling below, we warn the reader that this may not be valid in highly anisotropic manifolds (for example see~\cite{Collins:2022nux} for counter-examples where this assumption is not valid). Ideally, one would check the FL inequalities in a full construction where these assumptions are not needed but we leave this for the future.

Finally, we quote the $d$-dimensional Planck mass which can also be obtained from dimensional reduction of the Type II action:
\begin{align}
    \label{eq:ddimPlanck}
    M_d^{d-2} = \frac{M_s^8 W_{10-d}}{g_s^2}.
\end{align}
This will play a role in determining the electric field of a generic charged Nariai BH in $d$-dimensions which is $E \sim M_d^{\frac{d-2}{2}} H$ with $H$ the Hubble rate. 

\subsection{The FL bound in Type II}
With the above preliminaries out of the way, we are now ready to identify the coefficients $g$ that feature in our discussion of the FL bound in section \ref{sec:FLforBranes}. Later in this section we will discuss other couplings that show up in string theory, such as the analogue of $\gamma$ introduced in sec.~\ref{sec:OtherBraneCharges}, and argue why ignoring them is justified for this comparison. We will again split our discussion into cases where $p$ is even and odd, starting with the former. For the case of even $p$, we present a few more details to build intuition.

\subsubsection{Even $p$-branes}
\label{sec:FLcheckEvenp}
We want to consider the FL relation between the tension of a $p$-brane in $d$ dimensions and its coupling to gauge fields as in eq.~\eqref{eq:evenpAction}. However, if we start with the branes of Type II string theory, there are multiple ways to get a $p$-brane after compactification to a lower $d$-dimensional theory. For instance:
\begin{itemize}
    \item we can get a $p$-brane by considering an unwrapped D$p$-brane of the higher dimensional theory, or
    \item we can a get a $p$-brane by wrapping a D$(p+n)$-brane of the higher dimensional theory on a compact $n$-cycle of the internal manifold. 
\end{itemize}
In the second case, the $p$-brane in the non-compact dimensions can have various couplings that it can inherit from the different terms in eq.~\eqref{eq:CSaction}. This also means that we can get our desired coupling parameter in~\eqref{eq:evenpAction} in multiple ways. We will discuss both of these possibilities in turn.

For an unwrapped brane, the discussion is simple. The tension is as given in equation~\eqref{eq:pbraneTension} and all we need to do is identify the coupling $g$. Since we are assuming $p$ is even without wrapping any internal dimensions, our discussion is about Type IIA D-branes. For these branes, there is a coupling to $C_1$ that is exactly of the form shown in~\eqref{eq:evenpAction}. Rescaling the fields $C_1$ and $F^{\rm (B)}$ so that they have canonical kinetic terms gives:
\begin{align}
    \label{eq:gDbraneUnwrapped}
    g_p^{\rm unwrapped} = \left(\frac{g_s^{p/2}}{M_s^{6+p(p-3)/2} W_{10-d}}\right)^{1/2}
\end{align}
where again $W_{10-d}$ denotes the volume of the $10-d$ internal dimensions. Using this, together with the expression for the D$p$-brane tension in~\eqref{eq:pbraneTension} and the $d$-dimensional Planck mass~\eqref{eq:ddimPlanck}, it is easy to see that these branes satisfy the conditions in~\eqref{eq:FLevenp}. In fact, the Type IIA D-brane saturate the condition on the radius (i.e. the second inequality of~\eqref{eq:FLevenp}) and they automatically satisfy the condition on the tension (i.e. the first inequality). 

Let us demonstrate the above claim in the simple example of a D2-brane. For a D2-brane, we can simply read off the coupling from the CS action in Type IIA. This contains the term:
\begin{align}
I_{\rm D2} \supset M_s \int_{\rm WV} C_1 \wedge F^{\rm (B)}
\end{align}
which is identical to the coupling we assume for these branes. To get our coupling $g$, we rescale the fields $C_1$ and $F_B$ so they have canonical kinetic terms as we assumed in eq.~\eqref{eq:evenpAction}. After this rescaling, we find:
\begin{align}
g_{\rm D2} = \left(\frac{g_s}{M_s^{5} V}\right)^{1/2}
\end{align}
which agrees with the general expression in eq.~\eqref{eq:gDbraneUnwrapped} for $p=2$. We can use this to check the FL bound which we recall is given by:
\begin{align*}
T_{\rm D2} &\gtrsim \min \left[\left(g_{\rm D2} M_d^{\frac{d-2}{2}} H\right)^{6/5}, \left(g_{\rm D2} M_d^{\frac{d-2}{2}}\right)^2 \right] \\
&= \min \left[ \left(\frac{H^2 M_s^3}{g_s}\right)^{3/5}, \frac{M_s^3}{g_s} \right].
\end{align*}
It is easy to check the first quantity in the square brackets is smaller than the second as long as $g_s H^3 < M_s^3$ which is certainly the case for weak coupling and $H < M_s$. That said, the D2-brane tension saturates the second expression in the square brackets and therefore this bound is obeyed. The argument and the conclusion is similar for the cases $p=6$ and $p=8$. That is to say these branes saturate the condition derived from requiring $R_* \lesssim H^{-1}$ and satisfy the other inequality with room to spare at weak string coupling and for $H < M_s$.

We now turn to the case of wrapped branes. In a way reminiscient of the behaviour of the WGC under dimensional reduction~\cite{Heidenreich:2015nta}, we will see that the FL bound becomes monotonically weaker when applied to wrapped branes in compactifications. This happens in a non-trivial way as the relevant tensions and couplings change when considering wrapped branes. The two changes compensate to ensure that the bound continues to be satisfied. For wrapped branes, the coupling we are interested in may come from various different terms in the higher dimensional action and its coefficient will depend on its UV origin. As such, and to gain some intuition, we will start by working out two simple possibilities and then discuss the most general case.

The first possibility is that the coupling originates from a CS term in the higher dimensional theory that already includes $C_1$. As mentioned previously, we consider a D$(p+n)$-brane wrapping an internal $n$-cycle. In this context, the coupling we are looking for originates from a CS term of the form:
\begin{align}
    I \supset M_s \int_{(p+n)+1} C_1 \wedge \left(F^{\rm (B)}\right)^{p/2} \wedge \left(F^{\rm (B)}\right)^{n/2}
\end{align}
after integrating over $n$ internal directions. We assume that integrating over the $n$ internal directions does not produce parametrically large $F^{(B)}$ flux numbers. Also note that we need $n$ to be even and this term is consistent with the higher dimensional coupling found in Type IIA. After integrating over the $n$ internal directions, we need to ensure the kinetic term of the brane gauge field is canonical by rescaling $F^{\rm (B)}$ as in eq.~\eqref{eq:FBscaling} (of course we replace $p$ by $p+n$ in this expression). We also need to rescale the $C_1$ field as per eq.~\eqref{eq:RRfieldscaling1}. We can now read off the tension and coupling of this $p$-brane:
\begin{align}
    T_p = \frac{M_s^{p+1} \mathcal{V}_n}{g_s} \quad ; \quad g_p = \left(\frac{g_s^{p/2}}{M_s^{6+p(p-3)/2} W_{10-d}}\right)^{1/2} \frac{1}{\mathcal{V}_n^{p/4}},
\end{align}
where we write $\mathcal{V}_n \equiv M_s^n V_n$ for the volume of the cycle wrapped by the brane in string units. In the second equality, the expression in parentheses is exactly $g_p^{\rm unwrapped}$ from eq.~\eqref{eq:gDbraneUnwrapped}. Since $\mathcal{V}_n > 1$ by assumption, we learn that the tension of this brane is higher than the unwrapped brane and its coupling is lower than that of the unwrapped brane. For $p=2$, recall that FL bounds (a power of) the brane tension from below by a quantity proportional to $g$. This modification to the tension and coupling therefore only weakens the bound. For $p=6,8$, FL gives an upper bound on (a power of) the tension in terms of a quantity inversely proportional to $g$. For these two cases, we must check the powers of $\mathcal{V}_n$ more carefully. For the case $p=6$, FL gives:
\begin{align}
    T_6 \lesssim \max \left[
    \left(\frac{1}{g_6 M_d^{\frac{d-2}{2}} H}\right)^{14/5}, 
    \left(\frac{1}{g_6 M_d^{\frac{d-2}{2}}}\right)^2
    \right].
\end{align}
We already checked that $T_6$ saturates the second term in the square brackets which is in fact the smaller of the two so that the inequality is satisfied. In the case we have at hand, $T_6$ would be larger than that of the unwrapped brane by a factor of $\mathcal{V}_n$. On the other hand, the second term in the square bracket is larger by a factor of $\mathcal{V}_n^3 > \mathcal{V}_n$ so that the inequality only becomes easier to satisfy. For $p=8$, there is a similar story with different exponents and in that case the change in the tension exactly compensates the factor of $\mathcal{V}_n$ in the second term in square brackets. This again means that the inequality continues to be satisfied for wrapped branes if it was already satisfied for unwrapped branes as we indeed showed earlier.

The second simple possibility is that the coupling we are considering originates from a CS term with the correct number of $F_B$ fields but (necessarily) a higher form RR field. This means that we must reduce the form degree of the RR field when finding the coupling of the wrapped brane. Again we consider a D$(p+n)$-brane wrapping an $n$-cycle. This time, the coupling we are looking for originates from a CS term of the form:
\begin{align}
    I \supset M_s^{n+1} \int_{(p+n)+1} C_{n+1} \wedge \left(F^{\rm (B)}\right)^{p/2},
\end{align}
after integrating over the $n$ internal directions wrapped by the brane. In this case, $n$ may be even or odd so that we may find the above term in either Type IIA or Type IIB. Again, we have to rescale the fields appropriately to identify the relevant coupling. This gives:
\begin{align}
    T_p = \frac{M_s^{p+1} \mathcal{V}_n}{g_s} \quad ; \quad g_p = \left(\frac{g_s^{p/2}}{M_s^{6+p(p-3)/2} W_{10-d}}\right)^{1/2} \frac{1}{\mathcal{V}_n^{(p-4)/4}}.
\end{align}
We have again written $g_p^{\rm unwrapped}$ in parentheses showing that it is only modified by an overall factor. We now check whether this modification can lead to a violation of the FL inequalities in~\eqref{eq:FLevenp}. We start with $p=2$, where FL gives an upper bound on $T_2^{5/6}$ proportional to $g_2$. From the previous equation, we see that the tension and the coupling both increase for this wrapped brane. However, checking the powers of $\mathcal{V}_n$ it is easy to see that the FL bound continues to be satisfied for these wrapped branes if it is satisfied for the unwrapped ones. This is also the case for $p=6,8$.

Following the two examples we discussed above, we are now ready to discuss the most general case where we consider a coupling to a higher form RR field which can also contain higher powers of $F^{\rm (B)}$ (than the required $p/2$). In this case, the world volume action of the wrapped D$(p+n)$-brane has a coupling of the form:
\begin{align}
    I \supset M_s^{p+n+1-2q} \int_{(p+n)+1} C_{p+n+1-2q} \wedge \left(F^{\rm (B)}\right)^{p/2} \wedge \left(F^{\rm (B)}\right)^{q-p/2},
\end{align}
and we again need to integrate over $n$ internal directions to get the coupling we use to obtain the FL bound. This gives:
\begin{align}
    T_p = \frac{M_s^{p+1} \mathcal{V}_n}{g_s} \quad ; \quad g_p = \left(\frac{g_s^{p/2}}{M_s^{6+p(p-3)/2} W_{10-d}}\right)^{1/2} \frac{\mathcal{V}_{p+n-2q}}{\mathcal{V}_n^{p/4}}.
\end{align}
It is easy to see that this reduces to the simple cases above for particular choices of $2q = (p+n)$ and $2q = p$ respectively. In general, we have $ p \leq 2q \leq p +n$ which means that the coupling $g_p$ is related to the $g_p^{\rm unwrapped}$ by a factor that can be either larger or smaller than unity. Let us check how the FL inequality is satisfied for all three cases of even $p$ keeping in mind that the unwrapped branes satisfy the inequality.

\underline{${\bf p=2}$:}
In this case we want to verify:
\begin{align}
    \label{eq:T2LowerBound1}
    T_2 &\gtrsim \min \left[\left(g_2 M_d^{\frac{d-2}{2}} H\right)^{6/5}, \left(g_2 M_d^{\frac{d-2}{2}}\right)^2 \right].
\end{align}
Recall that the unwrapped branes saturate the second term in square brackets which is larger than the first (at weak coupling and $H< M_s$) and these branes therefore satisfy FL with room to spare. When considering wrapped branes, the LHS of this inequality increases by a factor of $\mathcal{V}_n$. The second term on the RHS changes by a factor:
\begin{align}
    \frac{\mathcal{V}^2_{n+2(1-q)}}{\mathcal{V}_n}.
\end{align}
For $q=1$, we recover the second of the simple cases we looked at earlier and the FL bound continues to be saturated. For $q>1$, the inequality is satisfied but no longer saturated since $\mathcal{V}_{n+2(1-q)} < \mathcal{V}_n$. This last statement makes an assumption about the isotropy of the wrapped cycle. It would be interesting to check how the story is modified when this assumption is not valid. That said, violating~\eqref{eq:T2LowerBound1} should involve a large hierarchy set by $H/M_d$ (to some power) and it is unclear to us if this can be achieved using anisotropy of an internal manifold. We leave this investigation for future work.

\underline{${\bf p=6}$:} In this case, the FL inequality reads
\begin{align}
    T_6 \lesssim \max \left[
    \left(\frac{1}{g_6 M_d^{\frac{d-2}{2}} H}\right)^{14/5}, 
    \left(\frac{1}{g_6 M_d^{\frac{d-2}{2}}}\right)^2
    \right].
\end{align}
Again, the tension of unwrapped D6-branes saturate the second term in square brackets. This is smaller than the first term so the inequality is easily satisfied. When considering wrapped branes, the LHS increases by a factor of $\mathcal{V}_n$. On the other hand, the second term on the RHS changes by a factor of:
\begin{align}
    \frac{\mathcal{V}_n^3}{\mathcal{V}^2_{n+2(3-q)}}.
\end{align}
For $q=3$, the change is again marginal. For $q>3$, it is easy to see that the RHS increases faster than the LHS for $\mathcal{V}_n > \mathcal{V}_{n+2(3-q)}$ (which again assumes isotropy). 

\underline{${\bf p=8}$:} The FL bound for these branes is:
\begin{align}
    T_8 \lesssim \max \left[
    \left(\frac{1}{g_8 M_d^{\frac{d-2}{2}} H}\right)^{9/8}, 
    \left(\frac{1}{g_8 M_d^{\frac{d-2}{2}}}\right)
    \right].
\end{align}
Here, the story is again very similar to the previous case. The only difference is that the second term on the RHS changes by a factor:
\begin{align}
    \frac{\mathcal{V}_n^2}{\mathcal{V}_{n+2(4-q)}},
\end{align}
which ends up giving exactly the same behaviour as the $p=6$ case. We have therefore shown that FL is satisfied for wrapped branes if it is already satisfied for unwrapped branes. 

\subsubsection{Odd $p$-branes}
\label{sec:FLcheckOddp}
We now turn to the case where $p$ is odd. This discussion is very similar to the one of the previous subsection and we follow the same logic. Namely, we consider first the simple cases followed by the most general scenario. Mirroring the discussion above, we would like to start by checking the FL bound for unwrapped branes. However, it is easy to see that this is an impossible task since the coupling in~\eqref{eq:oddpAction} does not arise directly for any of the D-branes in Type II. We must therefore consider wrapped branes throughout the discussion.

To get an odd $p$ brane with our coupling to bulk gauge fields, the simplest possibility is to wrap a D$(p+1)$-brane (of Type IIA) on a 1-cycle. This is simply the dimensional reduction of the unwrapped case considered in the even $p$ discussion. The relevant coupling in the higher dimensional theory is:
\begin{align}
    I \supset M_s \int_{(p+1)+1} C_1 \wedge \left(F^{\rm (B)}\right)^{(p+1)/2}.
\end{align}
When this brane is wrapped on a 1-cycle of radius $R$, we can perform one of the spatial integrals to get the term~\eqref{eq:oddpAction} in the dimensionally reduced action. The tension of the $p$-brane in the lower dimensional theory is just $T_{p+1}R$. As usual, to read off the relevant coupling we first rescale the fields so they have canonical kinetic terms. This gives:
\begin{align}
    \label{eq:oddpwrappedonCircle}
    g_p = \left(\frac{g_s^{\frac{p+1}{2}}}{M_s^{8+\frac{(p+1)(p-3)}{2} } W_{10-d}}\right)^{1/2}\left(M_s R\right)^{\frac{3-p}{4}}.
\end{align}
We can use this, along with the expression for the $d$-dimensional Planck mass and the brane tension to check the FL bound in eq.~\eqref{eq:FLoddp}. We find the same pattern that we got with $p$ even, which is that the FL inequality is satisfied by saturating the condition on the bubble radius. We illustrate this for $p=5$ since all other odd $p$ cases work in exactly the same way. For $p=5$, we start by recalling the FL inequality and substituting the expressions for the brane tension and coupling:
\begin{align}
    T_5 &\lesssim \max\left[\left(\frac{1}{g_5 M_d^{\frac{d-2}{2}}H}\right)^3, \left(\frac{1}{g_5 M_d^{\frac{d-2}{2}}}\right)^{2} \right] \\
    &= \max \left[
    \left(\frac{M_s}{H}\sqrt{\frac{M_s^5 R}{g_s}}\right)^{3},
    \frac{M_s^7 R}{g_s}
    \right].
\end{align}
It is clear that this tension saturates the bound coming from the bubble radius, i.e. the second quantity in square brackets. This quantity is smaller than first as long as $g_s (H/M_s)^6 < M_s R$ which is always the case in the weak coupling and large volume limit we are considering.

Next, we consider another simple case where the higher dimensional brane couples to a higher form RR field and the minimal number of brane world-volume gauge fields required to reproduce eq.~\eqref{eq:oddpAction}. This is analogous to the second of the simplified scenarios we discussed for even $p$. That said, we start with:
\begin{align}
    I \supset M_s^n \int_{(p+n)+1} C_{n} \wedge \left(F^{\rm (B)}\right)^{\frac{p+1}{2}}.
\end{align}
We will integrate over $n-1$ indices from $C_n$ and one index from a factor of $F^{\rm (B)}$ to get an axion. After rescaling, we get:
\begin{align}
    T_p = \frac{M_s^{p+1}\mathcal{V}_n}{g_s} \quad ; \quad g_p = \left(\frac{g_s^{\frac{p+1}{2}}}{M_s^{8+\frac{(p+1)(p-3)}{2} } W_{10-d}}\right)^{1/2} M_s R\frac{\mathcal{V}_{n-1}}{\mathcal{V}_n^{\frac{p+1}{4}}},
\end{align}
where $\mathcal{V}_{n-1}$ is the volume of the cycle used to reduce the form degree of $C_n$ in string units and $R$ is the radius of the circle we used to reduce the world volume 1-form field to an axion (called $\theta$ in eq.~\eqref{eq:oddpAction}). It is easy to check that FL is again satisfied for this tension and coupling. Let us illustrate this for the case $p=1$ where the FL bound is:
\begin{align}
    T_1 \gtrsim \min\left[
    g_1 M_d^{\frac{d-2}{2}} H
    , \left(g_1 M_d^{\frac{d-2}{2}}\right)^2
    \right].
\end{align}
We already saw that the minimally wrapped brane has a tension that saturates the second quantity on the RHS. For a wrapped brane, the LHS increases by a factor $\mathcal{V}_n$. On the other hand, the second quantity in square brackets changes by a factor:
\begin{align}
    \frac{\left(M_s R \mathcal{V}_{n-1}\right)^2}{\mathcal{V}_n}.
\end{align}
For an isotropic manifold, the above reduces to $\mathcal{V}_n$ and the FL bound remains marginal.

Finally, we proceed to discuss the most general scenario. Suppose we start with a D$(p+n)$-brane that wraps $n$ internal directions. We want to integrate over the $n$ internal directions to find a coupling of the form eq.~\eqref{eq:oddpAction}. In general the higher dimensional coupling has this form:
\begin{align}
    I \supset M_s^{n+p+1-2q} \int_{(p+n)+1} C_{n+p+1-2q} \wedge \left(F^{\rm (B)}\right)^q.
\end{align}
Integrating over the $n$ internal directions, we can reduce the form degree of the RR field $C$ and get the desired factors of the world volume gauge fields. The coupling is then identified after an appropriate rescaling of all fields and we find:
\begin{align}
    T_p = \frac{M_s^{p+1}\mathcal{V}_n}{g_s} \quad ; \quad g_p = \left(\frac{g_s^{\frac{p+1}{2}}}{M_s^{8+\frac{(p+1)(p-3)}{2} } W_{10-d}}\right)^{1/2} M_s R\frac{\mathcal{V}_{p+n-2q}}{\mathcal{V}_n^{\frac{p+1}{4}}}
\end{align}
This expression reduces to the previous one when $q = (p+n)/2$ and $n=1$. In this case we also get that $\mathcal{V}_n = M_s R$. By checking various cases, it is easy to get convinced that the FL bounds hold for these tensions and couplings as well.

\subsection{Back-reaction effects}\label{sec:stringbackreaction}
In this section, we show that, in the parametrically weak coupling regime, the effects of back-reaction from higher form bulk fields is negligible if in the initial configuration their associated field strengths were vanishing. This result can be directly imported from section \ref{sec:back-reaction}. As an explicit example, consider the nucleation of an even $p$-brane in $d$ dimensions where initially there is only a non-vanishing $dC_1\sim E$ field. Based upon the field rescaling of section \ref{stringpreliminaries}, the action is
\begin{align}
     I_{\mathrm{E}} \sim & T_p\int_{\mathrm{WV}} \star 1+ig_{(1)}\int_{\mathrm{WV}}  A_{(1)} \wedge \underbrace{F^{(\mathrm{B})} \wedge \ldots \wedge F^{(\mathrm{B})}}_{\text{$p/2$ factors}}+ ig_{(k)}\int_{\mathrm{WV}}  A_{(k)} \wedge \underbrace{F^{(\mathrm{B})} \wedge \ldots \wedge F^{(\mathrm{B})}}_{\text{$(p+1-k)/2$ factors}}  \\ &   +\int_{\mathrm{WV}}  F^{(\mathrm{B})} \wedge \star F^{(\mathrm{B})}  + \int_{\mathrm{Bulk}}dA_{(1)}\wedge \star dA_{(1)}+\int_{\mathrm{Bulk}}dA_{(k)}\wedge \star dA_{(k)}.\nonumber
\end{align}
for odd $3 \leq k < p+1$, where we renamed $C$ to $A$ after rescaling (see eq.~\eqref{eq:RRfieldscaling2}), $F^{(B)}$ has been rescaled according to eq.~\eqref{eq:FBscaling}, $T_p\sim M_s^{p+1}/g_s$, and likewise the couplings scale as:
\begin{align}
    g_{(1)}\propto g_{s}^{\frac{p}{4}}, 
\end{align}
and for general $k$, 
\begin{align}
    g_{(k)}\propto  g_s^{\frac{p+1-k}{4}}.
\end{align}
From section \ref{sec:back-reaction}, $\delta dA_{(k)}\propto g_{(k)}(g_{(1)}E)^{\frac{p-k+1}{4-p}}$. In the regime in which $g_s^{\frac{p}{4}}E$ is held fixed, with $g_s\rightarrow 0$ and $E\rightarrow \infty$, we see that $\delta dA_{(k)}$ goes to zero.  Disregarding other factors, the important point is the scaling behavior $g_{(k)}\sim g_s^{\frac{p+1-k}{4}}\rightarrow 0$ as $g_s \rightarrow 0$ (recall that $k<p + 1$ always holds, since the dimension of the RR form cannot exceed that of the world-volume of the brane itself) implies that if we take $dA_{(k)}$ (with $k\neq 1$) to be vanishing initially, then it will remain small in the weak coupling limit. It then follows that the back-reaction is under control and remains small. Just as we had argued in  section \ref{sec:back-reaction}, this development is pleasing since if we had found that $\delta dA_{(k)}\sim g_s^{-n}$, with $n>0$, then our solutions may no longer have been valid since additional back-reaction effects would have become dominant. Therefore, the calculations of this section continue to remain valid even in the presence of additional RR bulk fields as long as we are at sufficiently weak coupling. Note that we assume parametrically weak coupling here since there may also be large factors of internal volumes in the expression for $g_{(k)}$. This assertion is made clearer in the following section where we include the volume factors explicitly.

\subsection{RR Charges}\label{sec:rrcharges}

We have seen in the previous section that the dimensionally reduced action of a D$p$-brane can have multiple couplings that are similar to the one we used to derived the FL bound albeit involving RR forms of different degrees. Additionally, we argued that the effect of back-reaction from these terms is negligible at parametrically weak string coupling. We will now turn to a discussion of the RR charges of nucleated branes, which were not included above. We will see that, if these higher form fields survive in the low energy action and the D-branes remain charged under them, we will again be led to consider parametrically weak coupling for our calculations above to remain valid. 

In this section, we will quantify the contribution of RR charges. We will calculate $\gamma$ for stringy branes and check that, assuming a dS landscape, there is a regime where we can indeed apply the FL argument. As before, our first check will focus on unwrapped branes. In this case, the world-volume action contains
\begin{align}
    I \supset M_s^{p+1} \int_{p+1} C_{p+1}.
\end{align}
which describes the D-brane charge under a higher form RR gauge field. Canonically normalizing this field, we find that the coefficient $\gamma$ introduced in~\ref{sec:OtherBraneCharges} is:
\begin{align}
    \gamma_p^{\rm unwrapped} = \frac{M_s^{p-3}}{\sqrt{W_{10-d}}}.
\end{align}
The condition to ignore this contribution when evaluating the the Euclidean action is (see also section~\ref{sec:OtherBraneCharges}):
\begin{align}
    \gamma_p^2 R_*^{p+3 -d} &< T_p \nonumber\\
    \implies g_s \left(\frac{H}{M_s}\right)^{d-p-3} &< \mathcal{W}
\end{align}
where we have introduced $\mathcal{W} = W_{10-d} M_s^{10-d}$ which is the volume of the internal manifold in string units. This latter condition is automatically satisfied for branes with codimension higher than one. However, for codimension one branes, we have $p = d-2$ and $H/M_s$ appears with a negative power on the LHS. The inequality may still be satisfied by taking $g_s$ to be parametrically small but this condition is stronger than the ones we have encoutered earlier which were automatically satisfied at any weak coupling and large volume.

The constraint we found to ensure the validity of our calculation in the presence of RR charges becomes even stronger\footnote{This is in contrast to the bound itself which becomes weaker for wrapped branes.} if we were to consider wrapped branes. For instance, let us take a $(p+n)$-brane wrapped on an $n$-cycle. This is charged under $C_{p+n+1}$ and it is easy to check that the coefficient $\gamma$ of this term is:
\begin{align}
    \gamma_p^{\rm wrapped} = \frac{M_s^{p-3}}{\sqrt{W}} \mathcal{V}_n
\end{align}
which is larger than $\gamma_p^{\rm unwrapped}$ by a factor of $\mathcal{V}_n > 1$. A larger $\gamma$ makes it more difficult to neglect the contribution of the RR charge in the Euclidean action. In fact the condition becomes:
\begin{align}
    g_s \left(\frac{H}{M_s}\right)^{d-p-3} &< \frac{\mathcal{W}}{\mathcal{V}_n^2}
\end{align}
which is no longer necessarily satisfied, even for branes with higher codimension that one. Nonetheless, just as before, we can in principle satisfy this condition by taking $g_s$ arbitrarily small so as to overcome the volume factors on the RHS. In this sense, the arguments we made above in favor of FL and its generalizations apply in the parametrically weak coupling regime in the string theory context. 

Finally, we mention that there are other couplings (such as those involving $F_B$) that could lead to $\int C_{p+1}$ on the dimensionally reduced brane. These will come with additional powers of $g_s$ and the consistency condition we derive from them should be weaker. While it may be impossible to find dS solutions at parametrically weak coupling~\cite{Dine:1985he}, this is the regime in which we can trust the calculations we presented.

\section{Conclusion and outlook}
\label{sec:conclusion}

In this paper, we applied the logic of the FL conjecture to brane sources in de Sitter space, arguing that this naturally leads to new bounds on the tension of branes coupled to 1-form gauge fields via a Wess-Zumino/Chern-Simons term on their world-volume. These bounds are presented in equations~\eqref{eq:resultSummary1} and~\eqref{eq:resultSummary2}, which apply to branes with and without world-volume gauge fields respectively. In the former case, we discussed the effects of back-reaction and other brane charges and explained why these can be consistently neglected. In the latter case, we presented a loophole where a light axion can allow branes to evade our bound. Finally, for branes with world-volume gauge fields, we verified our inequalities using the properties of D-brane in Type II string theory and found that they are satisfied. 

Several promising directions remain open to further explore the FL bound. First, there are the unsettled topics discussed in section~\ref{sec:FLReview} which also pertain to the original FL proposal. For instance, it is imperative that we understand the importance of back-reaction as the electric field of the Nariai black hole is screened. This is highlighted in~\cite{Aalsma:2023mkz} where these effects are taken into account for the single-particle decay channel. It is important to carry out a study that includes all decay channels and the effects of particle--anti-particle annihilations. These detailed studies may also determine the prefactor in the FL bounds. For the original FL bound on charged particles, the prefactor was conjectured in~\cite{Montero:2021otb} by matching to the magnetic WGC in dS~\cite{Huang:2006hc}. The same prefactor was found in the calculations of~\cite{Aalsma:2023mkz}. An interesting task is to check whether this prefactor is affected by the inclusion of additional decay channels. 

Additionally, one can look for connections between the original FL proposal~\cite{Montero:2019ekk} and the new inequalities presented in this paper. A discussion of wrapped branes and their constraints under FL has already appeared in~\cite{Montero:2021otb}, yet it would be valuable to revisit this analysis in light of our bounds on the tension of branes. This will also provide consistency checks on FL under dimensional reduction, which could shed more light on the FL bound itself (see example~\cite{Heidenreich:2015nta} and~\cite{Rudelius:2021oaz} for the use of dimensional reduction in the context of the WGC and the dS conjecture~\cite{Obied:2018sgi}). There may also be connections between our bounds and the axionic FL bound proposed in~\cite{Guidetti:2022xct}. This work obtained a bound on the axion decay constant using the geometry of the internal manifold. Notably, this does not constrain the axion mass which raises the intriguing possibility of seeking such a constraint on axion masses. This is especially important given the plethora of experimental axion searches (see for example~\cite{AxionLimits}). We note that our constraint in eq.~\eqref{eq:resultSummary2} might have provided a bound on axion masses if not for the loophole presented in section~\ref{sec:loopholes}. 

In section~\ref{sec:stringtheory}, we checked our bounds against the properties of D-branes in string theory and found that they are all satisfied in the parametrically weak coupling regime. We note that there is a more general version of the calculations we performed, which includes all couplings (such as $\gamma$ and $g_{(k)}$) from the beginning, that will remain valid without having to take parametrically small string coupling. We leave this exercise, which produces more general and complicated bounds, for future work. The advantage of these more general bounds is that they can be compared directly to string theory branes without the need to assume arbitrarily small $g_s$ and are therefore more relevant for dS constructions which may not exist at parametrically weak coupling. 

We end this section by commenting on potential phenomenological applications of the bounds we derived in this work. As pointed out in~\cite{Montero:2021otb}, the FL bound suggests that dS constructions may be more complicated that previously thought. In particular, in the context of uplifts using anti-branes in a warped throat, the assumption that the warped throat is decoupled from bulk dynamics may lead to contradictions with the FL bound. For example, FL would forbid light charged matter arising from the bulk manifold away from the warped throat. Similar statements can be made about branes, although, as we saw in section~\ref{sec:stringtheory}, it is typically easier for wrapped branes to satisfy our bounds compared to unwrapped branes. Nonetheless, manifolds may be complicated and there could be scenarios that lead to violations of FL such as the one we describe now. For a wrapped brane, the coupling we rely on in eq.~\eqref{eq:evenpAction}, say, may come from the dimensional reduction of a higher form RR gauge field along the cycle that the brane wraps. The coefficient of the kinetic term of a 1-form gauge field obtained through this dimensional reduction procedure depends on the particular cycle the brane wraps. Canonical normalization of this kinetic term means that this coefficient affects the coupling that appears in eq.~\eqref{eq:resultSummary1}. Consider as an example the case of 2-branes in the non-compact dimensions where the FL bound we present is a lower bound on the brane tension. It is easy to see that the coupling becomes smaller if the 2-brane is obtained by wraping a higher dimension brane on a cycle that is the smallest representative in its homology class. If the brane instead wraps a meta-stable cycle that is not the smallest representative, the coupling may become smaller or larger and may thus violate the inequality we have in eq.~\eqref{eq:resultSummary1}. This would be an interesting, although potentially difficult, check to carry out. 

One could also envision applications of our bounds to early-universe physics. For instance, if we were to determine the Hubble rate during an epoch of inflation, we would obtain lower limits on the tension of 2-branes. Assuming we live in a Type II string theory, these bounds can be translated to bounds on a combination of the string scaled and the string coupling (that describe the tension of D2-branes) and this may explain why string theory was not found at the LHC for example! With the above considerations in mind, we end by highlighting the richness of the FL bound as a topic of exploration from theoretical and phenomenological perspectives.

\section*{Acknowledgements}
We thank Prateek Agrawal, Alek Bedroya, Junwu Huang, Mehrdad Mirbabayi, Miguel Montero, Rudin Petrossian-Byrne, Mario Reig, Houri Tarazi, Cumrun Vafa and Victoria Venken for useful discussions. SH is grateful to the Particle Theory group of the Physics Department of the University of Oxford for support. SH is a Junior Research Fellow of Christ Church, University of Oxford. The work of GO is supported by a Leverhulme Trust International Professorship grant number LIP-202-014. For the purpose of Open Access, the author has applied a CC BY public copyright license to any Author Accepted Manuscript version arising from this submission.

\appendix
\section{Horizon structure}\label{appA}
In section \ref{sec:FLReview}, the line element is 
\begin{equation}
    ds^2=-h(r)dt^2+h(r)^{-1}dr^2 +r^2 d\sigma^2+r^2 \sin^2 \sigma d\phi^2,
\end{equation}
with lapse function (here we take $d=4$ for concreteness):
\begin{equation}\label{lapse}
    h(r)=1-\frac{2G_{\mathrm{N}} M}{r}+\frac{G_{\mathrm{N}} Q^2}{r^2}-\frac{r^2}{l^2}, \; \mathrm{where}\; l^2=\frac{3}{\Lambda}.
\end{equation}
It is helpful to write $h(r)=-u(r)/(lr)^{2}$ where we have defined
\begin{equation}
    u(r)=r^4-l^2 r^2+2G_{\mathrm{N}} Ml^2 r -G_{\mathrm{N}} Q^2 l^2,
\end{equation}
which is a quartic polynomial in $r$. In particular, since the coefficient of the cubic term is vanishing, it follows that the sum of roots of $u(r)$ is vanishing. The polynomial $u(r)$ factorizes, so that for general values of parameters $(M,Q,l)$:
\begin{equation}\label{eqnroots}
    h(r)=-\frac{1}{(lr)^2}(r-\alpha)(r-\beta)(r-{\gamma})(r+\alpha +\beta+\gamma).
\end{equation}
The roots give the radial coordinates of the horizon, asssuming the root is a positive real number. If the first three roots are positive, then the final root is negative. We therefore see that we have at most three horizons corresponding to the three positive roots at which $h(r)$ vanishes. The outermost root with largest $r$ gives the cosmological horizon, the middle root gives the black hole outer horizon, and the smallest root gives the black hole inner horizon. We are most interested in extremal solutions, which are solutions that arise when at least two roots are degenerate.

First consider the case in which all three horizons are coincident. This is the ultracold point. Then call all the positive roots $\alpha$. We find
\begin{equation}
    h(r)=-\frac{1}{(lr)^2}(r-\alpha)^3(r+3\alpha ).
\end{equation}
Expanding the above, and equating with eq.~(\ref{lapse}) gives $\alpha=l/\sqrt{6}$. This parameter is the radial coordinate of the three coincident horizons.

Second, we will consider the theory of two degenerate roots for general values of $(M,Q,l)$, before classifying spacetimes of special interest that occur from particular choices of these parameters. We may express the degeneracy of two roots by taking $\gamma\rightarrow \beta$ in eq.~\eqref{eqnroots}, so that $\beta$ is the radial coordinate of the two coincident horizons. Then,
\begin{equation}
    h(r)=-\frac{1}{(lr)^2}(r-\alpha)(r-\beta)^2(r+\alpha +2\beta),\label{eqn:horizonroots2}
\end{equation}
which, upon expanding out and collecting terms in powers of $r$, simplifies to the expression
\begin{equation}
    h(r)=-\frac{1}{(lr)^2}\Big(r^4 -r^2 ((\alpha+\beta)^2+2\beta^2)+r(2\beta(\alpha+\beta)^2)-\alpha \beta^2(\alpha +2\beta)\Big).\label{eqn:horizonroots3}
\end{equation}
Comparing this result with eq.~(\ref{lapse}), we may eliminate $\alpha$ and obtain a parametric system of equations in $\beta$, 
\begin{equation}
    \frac{G_{\mathrm{N}} M}{l}=\frac{\beta}{l}\Bigg(1-2\Big(\frac{\beta}{l}\Big)^2\Bigg),\quad \frac{\sqrt{G_{\mathrm{N}}}Q}{l}=\frac{\beta}{l}\Bigg(1-3\Big(\frac{\beta}{l}\Big)^2\Bigg)^{1/2}.
\end{equation}
This is precisely the parametric plot Fig. \ref{fig:ParametricPlotNariai}. We see that $0 \leq \beta/l \leq 1/\sqrt{3}$ since we keep the charge real. Note that as $l\rightarrow\infty$ (i.e. $\Lambda\rightarrow 0$), we find that $\sqrt{G_{\mathrm{N}}}M=Q$, indicating a Reissner-Nordstr\"{o}m solution in Minkowski background, as expected. Furthermore, when $0\leq \beta/l < 1/\sqrt{6}$, we are on the upper branch, when $1/\sqrt{6}<\beta/l \leq 1/\sqrt{3}$, and when $\beta/l=1/\sqrt{6}$, we are at the ultracold point.

We are also interested in the near-horizon geometry of observers in extremal spacetime configurations for charged black holes in de Sitter in $(3+1)d$. Since $h(r)$ contains at least a double zero on either the upper or Nariai branch\footnote{And a triple zero at the ultracold point.}, the $h'(r)=0$ on the radial coordinate of the coincident horizons. Defining $x=r-u$, we may expand
\begin{equation}
    h(r)dt^2\approx \Big(dt\sqrt{h(u)}\Big)^2\Bigg(1+\frac{x^2}{2h(u)/h''(u)}+\cdots\Bigg),
\end{equation}
and 
\begin{equation}
    h(r)^{-1}dr^2\approx \Bigg(\frac{dr}{\sqrt{h(u)}}\Bigg)^2\Bigg(1+\frac{x^2}{2h(u)/h''(u)}+\cdots\Bigg)^{-1}.
\end{equation}
Defining new coordinates, 
\begin{equation}
    y=\frac{x}{\sqrt{h(u)}}, \ \mathrm{and,}\ \tau=t \sqrt{h(u)},
\end{equation}
we see that 
\begin{equation}
    h(r)dt^2\approx d\tau^2\Bigg(1+\frac{h''(u)}{2}y^2\Bigg) \ \mathrm{and,}\ \frac{dr^2}{h(r)} \approx dy^2 \Bigg(1+\frac{h''(u)}{2}y^2  \Bigg)^{-1},
\end{equation}
where we have truncated the expressions to quadratic powers of $y$. Taking $r=u$ to be the radial coordinate of the coincident horizons, we see that the sign of $h''(u)$ determines whether the near horizon geometry is $\mathrm{AdS}_2\times \mathrm{S}^2$ (on the upper branch)\footnote{This outcome is expected since the near horizon geometry of an extremal Reissner-Nordstr\"{o}m black hole is precisely $\mathrm{AdS}_2\times \mathrm{S}^2$, with a background electric field.} or $\mathrm{dS}_2\times \mathrm{S}^2$ on the Nariai branch, in the $\{\tau,x,\sigma,\phi\}$ coordinate system. 
It is merely a tedious algebraic exercise to check that this sign is positive on the upper branch and negative on the Nariai branch.

However, one thing is clear without any further calculation at all: since the ultracold point has a triple zero, $h''(u)=0$ there and so the near horizon geometry becomes $\mathrm{Mink}_2\times \mathrm{S}^2$. We see clearly now via explicit redefinition of coordinates that the singularities of the metric stemming from the zeros of $h(r)$ corresponding to horizons were entirely coordinate singularities, except the singularity at $r=0$, which is a physical singularity. One point to mention is that just as in extremal Reissner-Nordstr\"{o}m, in addition to the near horizon metric being $\mathrm{AdS}_2\times \mathrm{S}^2$, there was also a uniform background electric field in the radial direction, here to we have a uniform background electric field $E=Q/r^2$ where $r$ is the $S^2$ radius. Near the ultracold point, $E\sim \MPl H$.

The results above generalize to higher dimensions (for instance, see \cite{Tangherlini:1963bw, Dias:2004px, Myers:1986un,Astefanesei:2003gw}) which have a similar shark fin shaped (sub-)extremal region.

\section{Dyonic black holes}
\label{sec:appendixDyonicBHs}
In this section, we will describe the classical dyonic black holes solutions. These can be obtained by solving the set of coupled Einstein and Maxwell equations of motion. The solution is spherically symmetric so a metric ansatz of the form:
\begin{align}
	\label{eq:dyonicMetric}
	ds^2 = -U(r) dt^2 + \frac{dr^2}{U(r)} + r^ d\Omega^2
\end{align}
is sufficient\footnote{This is not the most general spherically symmetric ansatz but it is enough for the dyonic black hole solution.}. This solution is supported by an electromagnetic tensor with
\begin{align}
	\label{eq:EMtensorStatic}
	F_{rt} = E(r), \qquad F_{\sigma\phi} = B(r) r^2 \sin \sigma
\end{align}
where $\sigma$ is the polar angle coordinate. We can further express the electric and magnetic fields in terms of the black hole charges:
\begin{align}
\label{eq:QEQMdefinitions}
\tilde{Q}_E &= \frac{1}{4\pi}\int * F \\
\tilde{Q}_M &= \frac{1}{4\pi}\int F.
\end{align}
With these definitions, the blackening factor takes the form:
\begin{align}
U(r) = 1 - \frac{2 G \tilde{M}}{\tilde{r}} + \frac{4\pi G (\tilde{Q}_E^2 + \tilde{Q}_M^2)}{\tilde{r}^2} - \frac{\Lambda}{3}\tilde{r}^2
\end{align}
where $G = (8\pi \Mpl^2)^{-1}$ is Newton's constant and $\tilde{M}$ is the black hole mass. In addition, in terms of these charges, we have expressions for the electric and magnetic fields:
\begin{align}
\label{eq:EBfields}
E &= \frac{1}{\tilde{r}^2}\tilde{Q}_E \\
B &= \frac{1}{\tilde{r}^2}\tilde{Q}_M.
\end{align}

Finally, it will be convenient to work in units where the Hubble radius is unity (i.e. $\Lambda = 3$) so we rescale the mass and charge parameters to the dimensionless form:
\begin{align}
\label{eq:MQscaling}
M \equiv \sqrt{\frac{\Lambda}{3}} G \tilde{M}, \qquad Q_{E,M}^2 \equiv \frac{4\pi G \Lambda}{3} \tilde{Q}_{E,M}^2.
\end{align}
This black hole spacetime can be superextremal for certain choices of the parameters $M, Q_{E,M}$. We show the extremality region in Fig. \ref{fig:ParametricPlotNariai}.

We will now specialise to the Nariai branch and look for $dS_2 \times S^2$ solutions in the FRW patch, similar to~\cite{Frob:2014zka,Montero:2019ekk}. Taking an ansatz for the metric of the form:
\begin{align}
\label{eq:metric}
ds^2 = \frac{1}{\sqrt{\Lambda/3}\tilde{r}(\tilde{t})}(-d\tilde{t}^2 + a(\tilde{t})^2 d\tilde{x}^2) + \tilde{r}(\tilde{t})^2 d\Omega_2^2.
\end{align}
In the strict Nariai limit, we will find that $ \tilde{r}(\tilde{t})$ is a constant which describes the radius of the $S^2$ and the function $a(\tilde{t})$ is an exponential, as usual in $dS$ space. The above ansatz is valid close to the Nariai branch which is the only part of solution space that will be relevant for us. This spacetime is supported by an energy momentum tensor of the form:
\begin{align}
\label{eq:energymomentum}
T{^\mu}_{\nu} = \mathrm{diag}(-\tilde{\rho}, \tilde{p}_1, \tilde{p}_2, \tilde{p}_2)
\end{align}
where $\tilde{\rho}$ is the total energy density in the electromagnetic fields and $\tilde{p}_1$ and $\tilde{p}_2$ are the total pressures. We will define the equation of state parameters $\alpha$ and $\beta$ such that:
\begin{align}
\tilde{p}_1 = \alpha \tilde{\rho} \quad ; \quad \tilde{p}_2 = \beta \tilde{\rho}.
\end{align}
Finally, we will rescale the tilde coordinates by $\sqrt{\Lambda/3}$ so that they are dimensionless (e.g. $t = \sqrt{\Lambda/3} \tilde{t}$) and similarly for $\tilde{r}$. The energy density (and pressures) will be rescaled by $\rho = \tilde{\rho}/\Mpl^2 \Lambda$. In terms of these dimensionless variables, the Einstein equations become:
\begin{align}
\ddot{(r^2)} + \frac{1}{r} - 3 r + 3r \alpha \rho = 0 \label{eq:eom1} \\
2r \frac{\ddot{a}}{a} - \frac{1}{r^2} - 3-\frac32 [3(\alpha -1) - 2(\alpha + 2\beta-1)] \rho = 0 \label{eq:eom2}
\end{align}
These are the same equations obtained in the original FL paper~\cite{Montero:2019ekk}. Our analysis will be different only because of the detailed values of $\alpha, \beta$ and $\rho$. We need to supplement these equations by one describing the evolution of the energy density $\rho$. This can be obtained by using the continuity equation which reads:
\begin{align}
\label{eq:continuity}
\frac{\dot{\rho}}{\rho} = -\frac{\dot{a}}{a}(1+\alpha) - \frac{\dot{r}}{2r}(3-\alpha+4\beta).
\end{align}
Note that this equation holds even when $\alpha$ and $\beta$ are functions of time which is the case when the branes are being nucleated and screening the electromagnetic fields.

Let us now study the Nariai limit solution of these equations. As in the case of the electrically charged black hole, we expect our dyonic black hole to have spatially constant electromagnetic fields. In our coordinates, the electromagnetic fields $E$ and $B$ are described by the field strength tensor:
\begin{align}
\label{eq:EMtensorFRW}
F_{\tilde{r}\tilde{t}} = \frac{a}{\sqrt{\Lambda/3}\tilde{r}}E, \qquad F_{\sigma \phi} = B \tilde{r}^2 \sin \sigma
\end{align}
where $\sigma$ is the polar angle and we note the similarity with equation~\eqref{eq:EMtensorStatic}. We will also define and rescale the electric and magnetic charges as in equations~\eqref{eq:QEQMdefinitions} and~\eqref{eq:MQscaling}. We will fix these charges and solve for the remaining quantities in terms of them. In particular we will use equation~\eqref{eq:eom1} to find the $S^2$ radius. Plugging in the expression for the energy density, we find the radius:
\begin{align}
\label{eq:S2radius}
r^2 = \frac{1+\sqrt{1-12[Q_M^2 + Q_E^2]}}{6}.
\end{align}
Demanding that the square root is real implies a bound on the total charge $\sqrt{Q_M^2 + Q_E^2} < 1/\sqrt{12}$. 

Finally, we calculate the electric and magnetic fields of these dyonic black holes. We start by quoting the expressions for the integer quantised charges $\tilde{q}$:
\begin{align}
\tilde{q}_M &= 2 g \tilde{Q}_M \\
\tilde{q}_E &=  \frac{4\pi}{g} \tilde{Q}_E.
\end{align}
We can then check that the electric and magnetic fields of the dyonic black hole are, up to $\mathcal{O}(1)$ numbers:
\begin{align}
E &\sim \Mpl \sqrt{\Lambda} \label{eq:Efield} \\
B &\sim \Mpl \sqrt{\Lambda} \label{eq:Bfield}
\end{align}

\section{Dyonic Nariai black holes with axions}
\label{sec:appendixDyonicwAxions}
In the presence of an axion, we study the theory given by:
\begin{align}
    \mathcal{L}=\frac{M_{\mathrm{Pl}}^2}2(R-2\Lambda)-\frac14F^2-\frac12(\partial a)^2-m_a^2f^2\left(1-\cos\frac af\right)+\frac\alpha{8\pi}\frac afF_{\mu\nu}F_{\alpha\beta}\varepsilon^{\mu\nu\alpha\beta}.
\end{align}

The equations of motion of this theory are given by:
\begin{align}
\nabla_\sigma\left(-F^{\sigma\rho}+\frac\alpha{2\pi}\frac af\varepsilon^{\sigma\rho\alpha\beta}F_{\alpha\beta}\right)&=0\\
\nabla_{[\mu}F_{\alpha\beta]}&=0 \\
-\nabla^\mu\nabla_\mu a+m_a^2f\sin\left(\frac af\right)-\frac\alpha{8\pi f}F_{\mu\nu}F_{\alpha\beta}\varepsilon^{\mu\nu\alpha\beta}&=0\\
M_{\mathrm{Pl}}^2(R_{\mu\nu}-\frac12g_{\mu\nu}R+\Lambda g_{\mu\nu})&= \nonumber\\
\hspace{1cm}F_{\mu\alpha}{F_\nu}^\alpha-\frac14g_{\mu\nu}F^2+\partial_\mu a\partial_\nu a-g_{\mu\nu}\left[\frac12(\partial a)^2+m_a^2f^2\left(1-\cos\frac af\right)\right].
\end{align}

Let us now study the Nariai solutions of these equations. As in the case of the electrically charged Nariai black hole, we expect our dyonic black hole to have spatially constant electromagnetic fields. In our case, this will also imply a spatially constant axion field as parallel electric and magnetic fields act as classical sources of the axion. In our coordinates, the electromagnetic fields $E$ and $B$ are described by the field strength tensor:
\begin{align}
\label{eq:EMfields}
F_{\mu\nu} = \begin{pmatrix}
0 & -\frac{a}{\sqrt{\Lambda/3}\tilde{r}}E & 0 & 0\\
\frac{a}{\sqrt{\Lambda/3}\tilde{r}}E & 0 & 0 & 0\\
0 & 0 & 0 & B \tilde{r}^2 \sin \sigma \\
0 & 0 & -B \tilde{r}^2 \sin \sigma & 0
\end{pmatrix}
\end{align}
where $\sigma$ is the polar angle (since we are reserving the symbol $\theta$ for the dimensionless axion field). The value of the (constant) axion field is obtained by solving:
\begin{align}
\label{eq:thetavalue}
\sin \theta_* = \frac{\alpha}{\pi} \frac{E B}{m^2 f^2}.
\end{align}
Note in particular that solutions will have $\alpha E B \leq \pi m^2 f^2$ since $\sin \theta_* \leq 1$. In order to see how this comes about, it is useful to define the black hole charges (rescaled for convenience):
\begin{align}
\label{eq:charges}
\tilde{Q}_M &= \frac{1}{4\pi} \int F \\
\tilde{Q}_E &= \frac{1}{4\pi} \int *F + K \frac{\alpha \theta}{\pi} F \\
Q_{M,E} &= \sqrt{\frac{\Lambda}{6\Mpl^2}} \tilde{Q}_{M,E}.
\end{align}

In terms of these charges, we have expressions for the electric and magnetic fields (see also \cite{Campbell:1991rz, Balakin:2017nbg}):
\begin{align}
B &= \frac{1}{\tilde{r}^2}\tilde{Q}_M \\
E &= \frac{1}{\tilde{r}^2}\left(\tilde{Q}_E - \frac{\alpha \theta_*}{\pi} \tilde{Q}_M\right).
\end{align}
Fixing these charges, it is then easy to see that~\eqref{eq:thetavalue} always has a solution (see also \cite{Fischler:1983sc}). This is because we can always choose large values of $\theta_*$ so that the electric field is small enough which ensures that the right hand side of~\eqref{eq:thetavalue} is smaller than unity.

On the other hand, it is not always possible to find real solutions to equation~\eqref{eq:eom1} with constant $S^2$ radius. Plugging in the expression for the energy density, we find the radius:
\begin{align}
\label{eq:S2radiuswithAxion}
r^2 = \frac{1+\sqrt{1-12[Q_M^2 + (Q_E- \frac{\alpha}{\pi}\theta Q_M)^2][1 + X(1-\cos \theta)]}}{6[1 + X(1-\cos \theta)]}
\end{align}
where we have defined the dimensionless ratio $X \equiv m^2 f^2 / \Mpl^2 \Lambda$. Demanding that the square root is real implies a bound on the magnetic charge $|Q_M| < 1/\sqrt{12}$. When $Q_M = 0$, there is also a similar bound on the electric charge which is however relaxed with $Q_M \neq 0$ as we can choose a large axion value to reduce the electric field as above. 

Given the radius of the sphere, we can solve for the axion value $\theta_*$ by subtituting the expressions for the electric and magnetic fields into equation~\eqref{eq:thetavalue}. This gives:
\begin{align}
    \sin \theta_* &= \frac{2\alpha}{3\pi} \frac{\Mpl^2 \Lambda}{m^2 f^2}\frac{1}{r^4} Q_M \left(Q_E - \frac{\alpha \theta_*}{\pi} Q_M\right).
\end{align}
As we mentioned above, this equation will always have a solution because $\theta_*$ can be chosen so the right hand side is smaller than unity. In particular, as $m\rightarrow 0$, the right hand side would become large unless the expression in parentheses becomes proportional to $X = m^2 f^2/\alpha \Mpl^2 \Lambda$. In this small $m$ limit, we find that the electric and magnetic fields are:
\begin{align}
E &\sim \Mpl \sqrt{\Lambda} \left(\frac{m f}{g\Mpl H}\right)^2 \label{eq:Efield} \\
B &\sim \Mpl \sqrt{\Lambda} \label{eq:Bfield}
\end{align}
so that the electric field is parametrically smaller than $\Mpl \sqrt{\Lambda}$.

\bibliographystyle{JHEP}
\bibliography{refs.bib}

\end{document}